\documentclass[11pt]{elsarticle}

\usepackage{lineno,hyperref}
\modulolinenumbers[5]
\usepackage{float}
\usepackage{graphicx}
\usepackage{subfig}
\usepackage{mathrsfs}
\usepackage{graphics}
\usepackage{float}
\usepackage{amsmath}
\usepackage{amsfonts}
\usepackage{mathrsfs}
\usepackage{amsmath}

\usepackage{newtxtext} % use times new roman font style

\usepackage{optidef}
\usepackage[subnum]{cases}

\usepackage{optidef}
\usepackage{graphicx}
\usepackage{float}
\usepackage{graphicx}
\usepackage{subfig}
\usepackage{mathrsfs}
\usepackage{float}
\usepackage{amsmath}
\usepackage{amsfonts}
\usepackage{booktabs}
\usepackage{natbib}
\usepackage{bbm}
\usepackage{algorithm}
\usepackage[noend]{algpseudocode}
\usepackage{multirow}
\algdef{SE}[DOWHILE]{Do}{doWhile}{\algorithmicdo}[1]{\algorithmicwhile\ #1}

\usepackage[letterpaper, portrait, margin=1in]{geometry}
\usepackage{bbm}

\usepackage{mathtools}

\DeclarePairedDelimiter\floor{\lfloor}{\rfloor}

\DeclareMathOperator*{\argmin}{arg\,min}

\usepackage{xcolor}
\hypersetup{
    colorlinks,
    linkcolor={red!50!red},
    citecolor={blue!50!blue},
    urlcolor={blue!80!blue}
}

\usepackage{optidef}

\usepackage{amsthm}

\theoremstyle{definition}

\makeatletter
\newcommand*{\centerfloat}{%
  \parindent \z@
  \leftskip \z@ \@plus 1fil \@minus \marginparwidth
  \rightskip \leftskip
  \parfillskip \z@skip}
\makeatother

\usepackage{amssymb,amsmath,caption}
\usepackage[hang,flushmargin]{footmisc}
\newcommand{\algorithmfootnote}[2][\footnotesize]{
  \let\old@algocf@finish\@algocf@finish% Store algorithm finish macro
  \def\@algocf@finish{\old@algocf@finish% Update finish macro to insert "footnote"
    \leavevmode\rlap{\begin{minipage}{\linewidth}
    #1#2
    \end{minipage}}%
  }%
}

\usepackage{setspace}
\usepackage{natbib}
\usepackage{algorithm}
\usepackage[noend]{algpseudocode}
\algdef{SE}[DOWHILE]{Do}{doWhile}{\algorithmicdo}[1]{\algorithmicwhile\ #1}

\usepackage{appendix}

\usepackage{array}
\usepackage{threeparttable}
\usepackage{multirow}
\usepackage{rotating}
\usepackage{makecell}

\journal{Transportation}

\biboptions{authoryear}

\begin{document}
\begin{frontmatter}

\title{Impact of unplanned service disruptions on urban public transit systems}


\author[label1]{Baichuan Mo\corref{mycorrespondingauthor}}
\author[label2]{Max Y von Franque}
\author[label3]{Haris N. Koutsopoulos}
\author[label5]{John Attanucci}
\author[label4]{Jinhua Zhao}

\address[label1]{Department of Civil and Environmental Engineering, Massachusetts Institute of Technology, Cambridge, MA 02139}
\address[label2]{Department of Urban Studies and Planning, Biology, Massachusetts Institute of Technology, Cambridge, MA 02139}
\address[label3]{Department of Civil and Environmental Engineering, Northeastern University, Boston, MA 02115}
\address[label5]{Massachusetts Institute of Technology, Cambridge, MA 20139}
\address[label4]{Department of Urban Studies and Planning, Massachusetts Institute of Technology, Cambridge, MA 20139}

\cortext[mycorrespondingauthor]{Corresponding author}

\begin{abstract}

This paper proposes a general unplanned incident analysis framework for public transit systems from the supply and demand sides using automated fare collection (AFC) and automated vehicle location (AVL) data. Specifically, on the supply side, we propose an incident-based network redundancy index to analyze the network’s ability to provide alternative services under a specific rail disruption. The impacts on operations are analyzed through the headway changes. On the demand side, the analysis takes place at two levels: aggregate flows and individual response. We calculate the demand changes of different rail lines, rail stations, bus routes, and bus stops to better understand the passenger flow redistribution under incidents.  Individual behavior is analyzed using a binary logit model based on inferred passengers’ mode choices and socio-demographics using AFC data. The public transit system of the Chicago Transit Authority is used as a case study. Two rail disruption cases are analyzed, one with high network redundancy around the impacted stations and the other with low. Results show that the service frequency of the incident line was largely reduced (by around 30\%$\sim$70\%) during the incident time. Nearby rail lines with substitutional functions were also slightly affected. Passengers showed different behavioral responses in the two incident scenarios. In the low redundancy case, most of the passengers chose to use nearby buses to move, either to their destinations or to the nearby rail lines. In the high redundancy case, most of the passengers transferred directly to nearby lines. Corresponding policy implications and operating suggestions are discussed.

\end{abstract}


\end{frontmatter}

% \linenumbers
\section{Introduction}\label{intro}

Urban public transit systems play a crucial role in cities worldwide, transporting people to jobs, homes, outings, and a variety of other activities. Millions rely on urban transit systems to provide them with transportation. However, transit systems are susceptible to unplanned delays and service disruptions caused by equipment, weather, passengers, or other internal and external factors.

Mitigating the impact of unplanned service disruptions is an important task for urban transit agencies. For this reason, it is important to recognize how a transit system is affected by service disruptions. The analysis framework for incident impacts can be summarized in Table \ref{tab_tasks}. 
The two main dimensions of analysis, supply and demand, can further be broken down into ``network performance'' and ``service'' for supply analysis and ``passenger flow'' and ``individual behavior'' for demand analysis. 

\begin{table}[H]
\center
\caption{Analysis framework for incident impacts}
\begin{tabular}{@{}lll@{}}
\toprule
                        & Analysis tasks      & Description                                                        \\ \midrule
\multirow{2}{*}{Supply} & Network performance    & Indicators such as resilience, vulnerability, redundancy \\
                        & Service          & Changes in agency's operations (e.g., headway, routing)    \\ \midrule
\multirow{2}{*}{Demand} & Passenger flow      & Demand changes at different stations, lines                \\
                        & Individual behavior & Passengers' mode choices under incidents                         
    \\ \bottomrule
\end{tabular}
\label{tab_tasks}
\end{table}

The network performance analysis usually uses graph theory-based techniques to calculate indicators related to incidents \citep{berdica2002introduction,xu2015modeling, yin2016evaluating, zhang2018resiliency}, such as network resilience, vulnerability, redundancy, and a variety of other properties. The service analysis focuses on changes in an agency's operations during the incident period including headway, routing, staffing, and other operator-controlled factors designed to mitigate the incidents. From the demand point of view, passenger flow analysis investigates the demand changes at different stations, lines, or regions of a network, presenting passenger's choices and flow redistribution after service disruptions. The individual behavior analysis focuses on studying the individual's response (such as mode choices, waiting time tolerance) to the incident and its relationship to the individual's characteristics (e.g., travel histories, demographics) \citep{rahimi2020analysis, rahimi2019analysis}. Surveys are usually used for such studies.  

Previous research has used a variety of methods to analyze the impact of service disruptions. Of these methods, the three most common are graph theory-based, survey-based, and simulation-based. Graph theory-based methods usually derive resilience or vulnerability indicators based on the network topology \citep{yin2016evaluating, zhang2018resiliency,xu2015modeling, berdica2002introduction}. These methods are effective for understanding high-level network properties related to incidents. Survey-based methods investigate passenger behavior during and opinions about the incident \citep{curriemuir,murray2014behavioral,fukasawa2012provision,teng2015development,lin2018subway}. Passengers' individual-level behavior is analyzed and understood using econometric models. Simulation-based methods simulate passenger flows on the transit network under incident scenarios \citep{balakrishna2008simulation,suarez2005impacts,hong2018self}. These studies can output many metrics of interest such as vehicle load changes, additional travel delays caused by incidents, distribution of the impact, etc. 

Recently, automated data collection systems in transit networks enable a data-driven analysis of the impacts of service disruptions. The two major sources are automatic fare collection (AFC) and automatic vehicle location (AVL) data. AFC data is collected when passengers tap their transit cards on smart card readers (in buses or rail station gates). The records include times, locations and card IDs. Depending on whether the fare system requires passengers to tap out, AFC data may only include tap-in records or both tap-in and tap-out records. AVL data records vehicle's (bus and train) time-dependent locations based on GPS and train tracking systems. From the AVL records, information such as headways can be inferred. Recently, a limited number of studies have been conducted using AFC and AVL data to look at unplanned transit disruptions. For example, \citet{sun2016estimating} analyzed three types of abnormal passenger flows during unplanned rail disruptions using AFC data with both tap-in and tap-out records. \citet{tian2018using} proposed a classification model to predict whether commuters switch from rail to other transportation modes because of unexpected travel delays using six months of AFC data. 

However, despite numerous studies on incident analysis, there are still research gaps. First, for the graph theory-based approaches, the network indicators such as redundancy are usually defined for the whole network, an OD pair, or a link, and do not consider the influence of the disruption duration. Incidents usually cause service interruptions at multiple links depending on the power system and rail track configuration. And the duration of an incident can vary from 5 minutes to several hours, resulting in various impacts on the network. An incident-based indicator that reflects the network's redundancy under an actual incident with a specific location and duration is needed. Second, the studies that leverage AFC data to analyze passengers' mode choices under disruptions are very limited. Such approachs would require the inference of both individual choices and socio-demographic information from the AFC data. Third, most of the previous studies on incident analysis only addressed one or two aspects in Table \ref{tab_tasks} using case studies of a single incident. A comprehensive study that analyzes all four dimensions of the problem with comparable case studies using AFC and AVL data is missing from the literature.

The paper aims to fill these research gaps by developing a data-driven methodology for the comprehensive analysis of the impact of unplanned rail disruptions on passengers and operations. Specifically, on the supply side, we propose an incident-based network redundancy index to analyze the ability of bus and rail networks to provide alternative services under a specific rail disruption. The impacts on operations are evaluated through headway changes across the systems. On the demand side, we calculate the demand changes at different rail lines, rail stations, bus routes, and bus stops to better understand the passenger flow redistribution under incidents. Individual behavior is analyzed using a binary logit model based on inferred passengers' mode choices and socio-demographics using AFC data. The public transit system of the Chicago Transit Authority (CTA) is used for a case study with two rail disruptions, one of which has high network redundancy and the other low. 

The main contributions of this paper are as follows:
\begin{itemize}
\item Propose an incident-based network redundancy index to reflect the system's ability to provide alternative services considering the integrated bus and rail systems. The index leverages the proposed concept of path throughput to incorporate the impact of the incident duration on the redundancy calculation.
\item Develop an incident analysis framework using AFC and AVL data and apply it to incidents with different characteristics. Specifically, we analyze two types of incidents with high and low redundancy separately from both demand and supply perspectives. 
\item Propose an individual mode choice analysis method using AFC data. The approach includes a travel mode inference model and a passenger demographics extraction model. To the best of our knowledge, this is the first study that adopts AFC data for individual mode choice analysis during incidents. 
\item Conduct an empirical study to demonstrate the proposed framework using AFC and AVL data from two real-world incidents in the CTA system. The corresponding policy implications and operation suggestions are also discussed. 
\end{itemize}

The remainder of this paper is organized as follows. Section \ref{liter} reviews the literature. Section \ref{sec_method} presents the methodology used in this study. Case studies and data are described in Section \ref{sec_case} and results are discussed in Section \ref{sec_result}. Section \ref{sec_conclude} concludes the paper and discusses the policy implications.

\section{Literature review}\label{liter}
There are generally four methods researchers use to analyze the impact of disrupted operations: graph theory-based, survey-based, AFC data-based, and simulation-based. Graph theory-based analysis is majorly used for supply network performance and supply service analysis. Survey and AFC data-based methods are primarily used for passenger flows and individual behavior analysis. Lastly, simulation-based analysis can be used for both supply and demand analysis. Each method has strengths and weaknesses depending on the context. 

\subsection{Supply analysis}
\subsubsection{Network performance}
Network performance analysis usually uses graph theory-based techniques to identify key aspects of the network's properties related to incidents, such as resilience, redundancy, and vulnerability based on graph theory (or complex network theory). For example, \citet{yin2016evaluating} studied subway networks with respect to disruptions, finding the weakness or critical locations of the network using ``network betweenness'' and ``global efficiency'' metrics. Similarly, \citet{zhang2018resiliency} built a general framework to assess the resilience of large and complex metro networks by quantitatively analyzing their vulnerability and recovery time using graph theory-based definitions.

Simulation is also used to evaluate the impact of incidents on networks. Usually, different hypothetical incident scenarios are tested. System performance metrics, such as travel delays and vehicle loads, are output to analyze the incident effects. For example, \citet{suarez2005impacts} looked at the effects of climate change on Boston's transportation system performance using a simulation model, suggesting almost a doubling in delays and lost transit trips due to a variety of climate change effects.

Redundancy is an important indicators for analyzing the network performance under incidents. Redundancy is best defined as ``the extent to which elements, systems, or other units of analysis exist that are substitutable, i.e., capable of satisfying functional requirements in the event of a disruption, degradation, or loss of function'' \citep{bruneau2003framework}. Redundancy has been widely studied, not just for transportation networks, but also in other areas including reliability engineering \citep{o2012practical}, communications \citep{wheeler1999network}, water distribution systems \citep{kalungi2003redundancy}, and supply chain and logistics \citep{sheffi2005supply}. Transportation-specific resiliency and redundancy studies include \citet{berdica2002introduction}, who developed a qualitative framework and basic concepts for vulnerability, resilience, and redundancy. Other studies, like \citet{wilson2006assessment}, \citet{murray2006comparison}, and \citet{goodchild2009development}, defined redundancy in the context of a specific transportation area of application. \citet{jenelius2015value} look at redundancy during line extensions. However, nearly all previous studies defined redundancy for networks, links, and OD pairs. This study proposes an incident-based redundancy index to evaluate the network's ability to satisfy functional requirements under a specific incident. Both incident location and duration impact the redundancy. Moreover, the bus system, which is an important alternative for rail but rarely considered in previous studies, is included in the redundancy calculation.

\subsubsection{Service Analysis}
Service analysis mainly focuses on changes in an agency's operations during an incident period. This type of analysis looks at headways, routing, staffing, shuttle services, and other operator-controlled factors designed to mitigate the incident. For example, \citet{nash2004railroad} developed a simulation model to analyze service variables such as headways and routing in the wake of disruptions. \citet{schmocker2005metro} evaluated different operating strategies in six metro systems under service disruptions. Service delays and recovering times are treated as performance indicators. Similarly, \citet{mo2020capacity} proposed an event-based simulation model that is capable of analyzing the impacts of incidents on service performance (e.g., headways).

\subsection{Demand analysis}
\subsubsection{Passenger flow}
Passenger flow analysis focuses on understanding how passengers choose alternative services \textbf{at an aggregated level}. Simulation-based methods can be applied to passenger flow analysis. For example, \citet{hong2018self} simulated passenger flows in a metro station during an emergency. Using AFC data, \citet{sun2016estimating} quantified three types of passenger flows: leaving the system, taking a detour, and continuing the journey but being delayed. This model was applied to the Beijing metro network. \citet{tian2018using} looked at unexpected train delay effects on Singapore’s MTR customers. Using AFC data, they built a classification model to predict whether commuters switch from MRT to other transportation modes because of unexpected train delays. \citet{wu2020data} used AFC data to detect passenger flow volumes and travel time increases under station closures. \citet{liu2021unplanned} uses AFC data to comprehensively analyze unplanned disruption impacts, especially on passenger flows with trip cancellation, station changes, etc.
 
\subsubsection{Individual behavior}
Individual behavior analysis usually focuses on individual responses, like mode choice, waiting time tolerance, and a variety of other variables. These studies are usually conducted using surveys. Surveys are a good means to understand individual choices. Revealed preference (RP) and stated preference (SP) are two major types of survey design. Examples of transit-oriented RP studies include \citet{curriemuir}, who conducted an RP survey to understand rail passengers’ behavior, perceptions, and priorities in response to unplanned urban rail disruptions in Melbourne, Australia. \citet{murray2014behavioral} used a web-based RP survey to understand the long-term impacts of a deadly metro rail collision in Washington DC. \citet{tsuchiya2008route} conducted an RP survey in Japan that looked at passenger choices of four alternative routes. \citet{pnevmatikou2011demand} used RP survey data to analyze the effect of a pre-announced closure of an Athens Metro Line. SP survey studies include \citet{kamaruddin2012public}, who studied the modal shift behavior of rail users after incidents. \citet{fukasawa2012provision} investigated the effect of providing information such as estimated arrival time, arrival order, and congestion level on passengers’ modal shift behavior in response to an unplanned transit disruption. A similar research was conducted by \citet{bai2014modeling}, who found that various socioeconomic attributes and experience with the systems had strong influences on travelers’ behavioral responses in the context of real-time information. Additionally, \citet{rahimi2019analysis, rahimi2020analysis} used an failure time model and a discrete choice model to analyze individuals' waiting time tolerances and mode choices, respectively, during unplanned service disruptions in Chicago using survey data. 

The major drawback of survey-based methods is that they are time-consuming and labor-intensive. Hence, it is important to develop individual behavior analysis methods using AFC data as an alternative. 

\subsection{Comparison between our study and the literature}
Table \ref{tab_liter} summarizes the various studies in the literature from three aspects: study methods, data sources, and research focus. The main methodologies include graph theory-based methods (GTB), simulation-based (SB), optimization models (OM), descriptive analysis (DA), statistical inference (SI), machine learning (ML), and econometric models (EM).  
\begin{table}[H]
\center
\caption{Summary of literature review}
\resizebox{1\linewidth}{!}{
\begin{tabular}{@{}llll@{}}
\toprule
\textbf{Study}                 & \textbf{Study Method}                          & \textbf{Data Sources}           & \textbf{Research Focus}                                              \\ \midrule
\citet{yin2016evaluating}         & GTB & Network, AFC  & $^1$NP - Efficiency                              \\
\citet{zhang2018resiliency}       & GTB  & Network   & NP - Resilience                                    \\
\citet{balakrishna2008simulation} & SB    & Network, Survey               & NP - Efficiency, Passenger flow, Service             \\
\citet{hong2018self}              & SB                               & Synthetic              & Passenger Flow                                             \\
\citet{suarez2005impacts}         & SB                              & Network, Geographical & NP - Resilience                                     \\
\citet{mo2020capacity}            & SB, OM                 & Network, AFC                             & Passenger Flow, Service                                             \\
\citet{jenelius2015value}         & SB          & Network, AFC          & NP - Redundancy                                     \\
\citet{adnan2017evaluating}       & SB                & Survey                          & NP - Efficiency, Service                                                   \\
\citet{schmocker2005metro}        & DA                          & AVL, AFC, Survey                & NP - Resilience, Service                                  \\
\citet{sun2016estimating}         & DA, SI    & AFC                             & Passenger Flow                                             \\
\citet{tian2018using}             & ML                           & AFC                             & Delay                                               \\
\citet{wu2020data}                & SI           & AFC                             & Passenger Flow                                               \\
\citet{liu2021unplanned}          & DA                           & AFC, AVL, Network               & Passenger Flow, Delay                                             \\
\citet{curriemuir}                & EM                              & Survey                          & Travel mode choice, User satisfaction                              \\
\citet{murray2014behavioral}      & EM                              & Survey                          & Travel mode choice                                          \\
\citet{tsuchiya2008route}         & EM                            & Survey                          & Travel mode choice                                   \\
\citet{pnevmatikou2011demand}     &EM                      & Survey                          & Travel mode choice                                  \\
\citet{kamaruddin2012public}      & EM                            & Survey                          &  Travel mode choice                                              \\
\citet{fukasawa2012provision}     & EM                           & Survey                          & Travel mode choice                                                   \\
\citet{bai2014modeling}           & EM                           & Survey                          & Travel mode choice                                                \\
\citet{lin2018subway}             & EM                             & Survey                          &Travel mode choice                                             \\
\citet{pnevmatikou2015metro}      &EM                           & Survey                          &Travel mode choice                                                \\
\citet{rahimi2019analysis}        & EM                     & Survey                          & Travel mode choice                                                    \\
\citet{rahimi2020analysis}        & EM                         & Survey                          & User waiting behavior                                               \\\midrule
\textbf{Current study}                  & GTB, EM, DA & AFC, AVL, Network               & \begin{tabular}[c]{@{}l@{}}Travel mode choice, Passenger flow, \\  NP - Redundancy, Service\end{tabular} \\ \bottomrule
\multicolumn{4}{l}{\begin{tabular}[c]{@{}l@{}}$^1$NP: Network performance.
\end{tabular}}
\end{tabular}
}
\label{tab_liter}
\end{table}

Our study presents a comprehensive analysis focusing on four aspects: travel mode choice, passenger flow, redundancy, and service. It is also exclusive based on AFC and AVL data.

\section{Methodology}\label{sec_method}

In this section, we present the building blocks and methods used to support the analysis framework of an unplanned incident. On the supply side, a method to calculate the network redundancy index under a certain incident is proposed, which reflects the network's ability to provide alternative routes when incidents occur. To analyze the agency's service, we calculate the headway distribution using AVL data. On the demand side, we describe how to analyze passenger flows under incidents using AFC data, and how to use AFC data to analyze passengers' mode choice using a binary logit model. 

To infer the effect of an incident, we compare data from the incident day to corresponding data from normal days. A ``normal day'' is defined as a recent day with the same day of the week and there are no incidents occurring in the incident line or nearby region during the incident period on that day. For example, if an incident happened on Friday 9:00-10:00 AM at a station, normal days can be all Fridays in recent months without incidents from 8:00-11:00 AM (a buffer is added to ensure normal services in a normal day) on the same line. Headways and passenger flow during the incident day are compared to those of normal days to reveal their difference. 

\subsection{Supply analysis}

\subsubsection{Network redundancy under incidents}

As mentioned in Section \ref{liter}, since redundancy is used to evaluate a network's functional response in the event of disruptions, it is important to develop an incident-specific redundancy index (opposite to network, link, or OD pair-specific in the literature). For a given incident, such an index can be used to evaluate the network's ability to provide alternative services under this incident. Furthermore, given the substitutional relationship between bus and urban rail systems, the proposed redundancy index in this study also explicitly considers the complementary role of bus and rail systems during the incident. 

Redundancy is usually a function of the number of available paths for each OD pair because more available paths correspond to more opportunities of realizing the impacted trips when encountering service disruptions  \citep{xu2015modeling}. Hence, network redundancy under incidents (NRUI) should capture the transport capacity of alternative paths during the incident. Typical path capacity is defined as the maximum number of passengers transported per time unit (i.e. service frequency times the vehicle capacity). It is a time-insensitive value, which means the travel times of paths are not considered. However, for the redundancy calculation, path travel times are also important because passengers may not successfully finish their trips during the incident period if they choose paths with a long travel time. This means that the time-insensitive path capacity does not reflect the actual ability of paths to move passengers. Hence, a time-sensitive path capacity should be used for the calculation. In this study, we propose a new metric for the calculation of NRUI. The basis of the analysis uses the concept of throughput, instead of the classic definition of path capacity. In our approach, throughput explicitly takes into account the travel time on each alternative path. Throughput is defined as the number of ``equivalent'' passenger trips that have been completed per time unit during the incident. If a passenger has completed half of the trip on an alternative path by the time an incident is over, the ``equivalent'' trip count is 0.5. 

\begin{figure}[H]
\centering
\subfloat[$D_I < L_p$]{\includegraphics[width=0.5\textwidth]{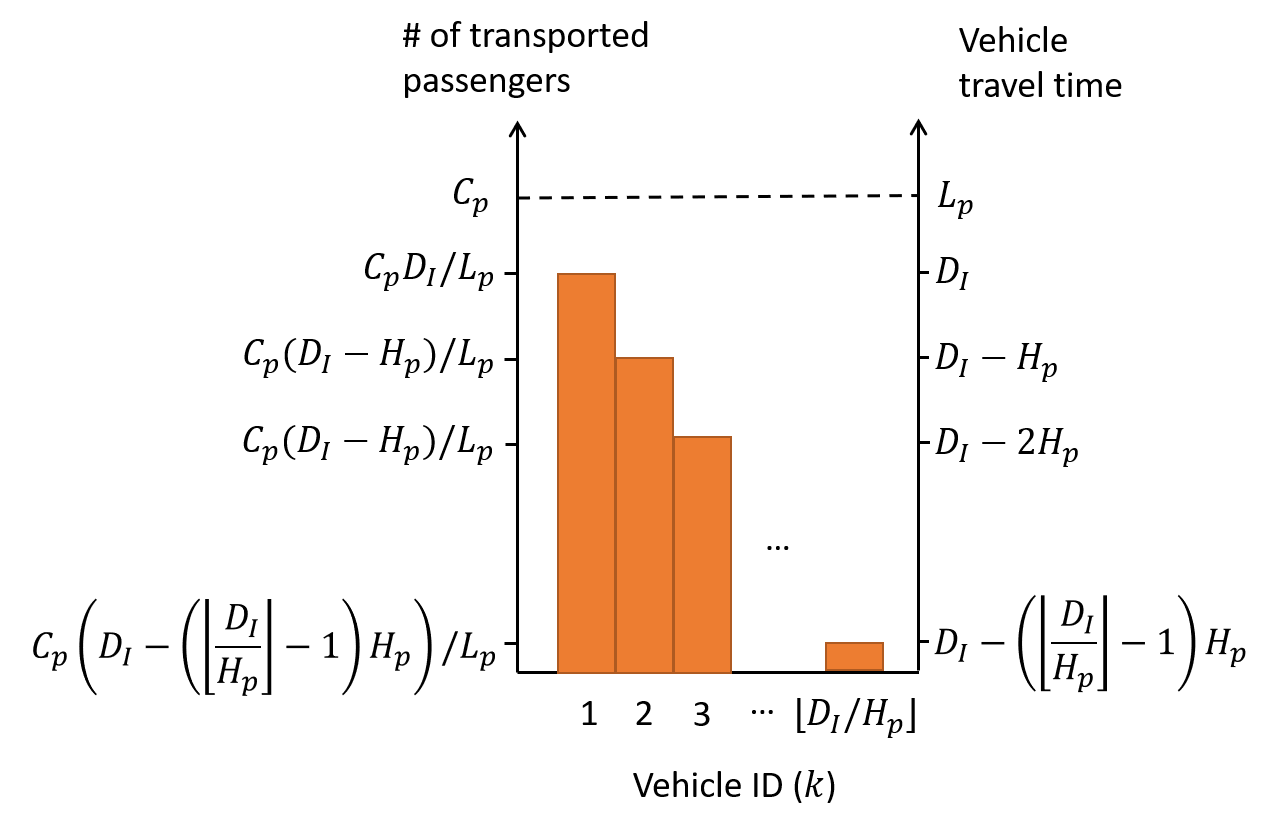}\label{fig_Ap_small_DI}}
\hfil
\subfloat[$D_I \geq L_p$]{\includegraphics[width=0.5\textwidth]{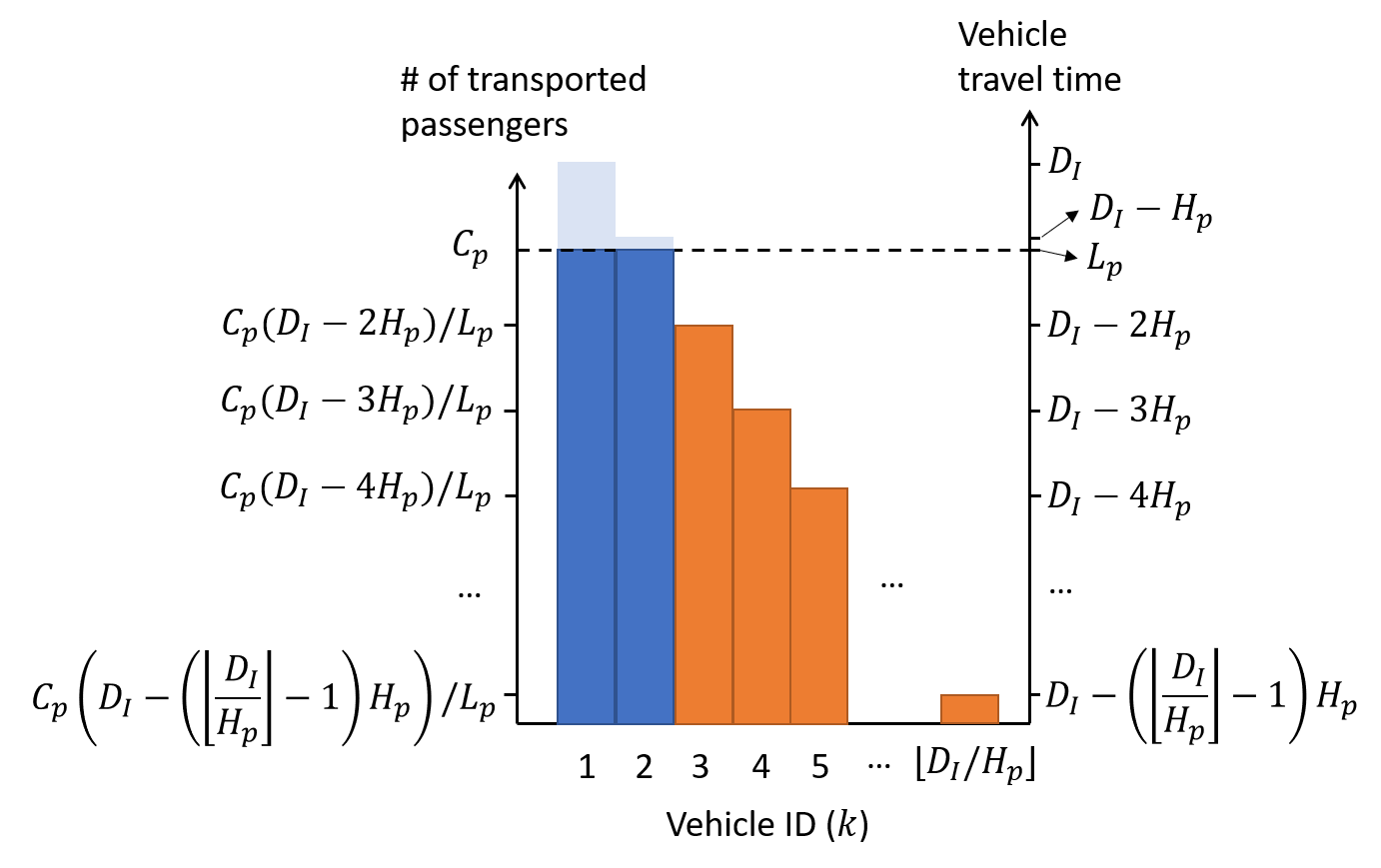}\label{fig_Ap_big_DI}}
\caption{Illustration of path throughput. The bars show the number of equivalent completed passenger trips during the incident period (unfinished trips are counted proportionally based on their travel time). The orange (blue) bars represent vehicles that cannot (can) finish the trips.}
\label{fig_Ap_illustration}
\end{figure}

More specifically, let $\mathcal{W}$ be the set of all OD pairs of the rail network. For an OD pair $w \in \mathcal{W}$, let $\mathcal{P}_w$ be the set of available paths for $w$ \emph{before the incident}. As we consider both bus and urban rail systems, a path $p \in \mathcal{P}_w$ may include segments of bus trips. $\mathcal{P}_w$ can be obtained in several ways, such as route choice surveys, Google Map API, and $k$-shortest paths. In this study, $k$-shortest path is used to obtain $\mathcal{P}_w$ with additional manual tuning to filter out unrealistic paths (e.g., too many transfers). Let $D_I$ be the duration of incident $I$, $H_p$ the headway of path $p$ (defined as the maximum headway of each segment of path $p$), $C_p$ be the vehicle (i.e. train or bus) capacity of path $p$ (defined as the minimum vehicle capacity over all segments of path $p$), and $L_p$ the travel time of path $p$. Then $\floor{D_I/H_p}$ is the total number of vehicles dispatched on path $p$ during the incident period. The throughput aims to capture the number of passengers at various stages of completing their trips \emph{during the incident}. Figure \ref{fig_Ap_illustration} illustrates how the equivalent number of passengers completing trips are calculated during the incident period. If $D_I < L_p$ (Figure \ref{fig_Ap_small_DI}), all vehicles in the path cannot reach the final destination. Therefore, the number of transported passengers is counted proportionally based on their travel time in the vehicle. For example, the first vehicle has traveled for $D_I$ during the incident period (i.e. $\frac{D_I}{L_p}$ of the total path length). We assume this is equivalent to $C_p\frac{D_I}{L_p}$ completed passenger trips. And it is easy to show that the $k$-th vehicle's travel time is $D_I - (k-1)H_p$, which corresponds to $C_p\frac{D_I - (k-1)H_p}{L_p}$ equivalent completed passenger trips during the incident period. If $D_I \geq L_p$ (Figure \ref{fig_Ap_big_DI}), the first vehicle can reach the destination. So it accounts for $C_p$ completed passenger trips. In the example shown in Figure \ref{fig_Ap_big_DI}, the second vehicle can also reach the destination during the incident (accounting for $C_p$ passenger trips), while the third cannot (accounting for $C_p\frac{D_I - 2H_p}{L_p}$ passenger trips). Therefore, combining these two scenarios, the number of equivalent completed passenger trips for vehicle $k$ can be calculated as  $\frac{\min{\{D_I - (k-1)H_p, L_p\}}}{L_p} \cdot C_p$.

Let $A_p$ be the throughput of path $p$ under incident $I$. From the analysis above, it can be formulated as 
\begin{align}
A_{p} = \frac{1}{D_I}\sum_{k=1}^{\floor{D_I/H_p}} \frac{\min{\{D_I - (k-1)H_p, L_p\}}}{L_p } \cdot C_p
\label{eq_Ap}
\end{align}
Eq. \ref{eq_Ap} counts the total number of equivalent passenger trips along path $p$  that have been completed per time unit during the incident (passengers who did not finish their trips are counted proportionally).

A larger value of $D_I$ implies that $A_p$ is less sensitive to the path travel time. On the extreme situation where $D_I \to \infty$, $A_p \to \frac{C_p}{H_p}$ (proof in \ref{append_cap}), which corresponds to the typical definition of capacity where $L_p$ does not matter. The intuition behind this is that the proposed $A_p$ limits the capacity calculation \emph{in the incident period}. When $D_I$ is large, even if passengers have a longer travel time on a path, the majority of the passengers impacted by the incident will have their trips completed. So the travel time is not very. On the contrary, if $D_I$ is small, most of the passengers using paths with long travel times cannot finish their trips. The typical definition of $\frac{C_p}{H_p}$ does not capture this important aspect. Hence, considering travel time in the redundancy calculation is much more representative of the actual conditions.  

In summary, $A_p$ is an indicator reflecting a path's ability to serve impacted trips during the incident period. And $\sum_{p\in\mathcal{P}_w} A_p$ reflects the ability of the network to provide services for OD pair $w$. 
Actually, one of the network-level definitions of redundancy in the literature is $\sum_{w\in\mathcal{W}}\sum_{p\in\mathcal{P}_w} A_p$ (where $A_p$ is defined differently), which measures the total path capacity in the network \citep{leistritz2013network}. 

In this study, we want to capture the incident-specific characteristics of redundancy. Let $\mathcal{W}_I$ be the set of all OD pairs with at least one path blocked due to incident $I$. Mathematically, $\mathcal{W}_I = \{$ $w\in \mathcal{W}: \exists p \in \mathcal{P}_w$ s.t. $p$ is blocked due to incident $I$ $\}$. Then, only passengers with OD in $\mathcal{W}_I$ are affected by the incident. Let the total path throughput of $w$ before the incident be $T_w$. 
\begin{align}
    T_w = \sum_{p\in {\mathcal{P}}_w} A_p  \quad \quad \forall w \in \mathcal{W}_I
\label{eq_Tw}
\end{align}

Because of the incident, passengers may augment their typical path choice alternatives with paths that were not considered before the incident. Hence, we define $\Tilde{\mathcal{P}}_w$ as the set of available paths for $w \in \mathcal{W}_I$ \emph{during the incident}. $\Tilde{\mathcal{P}}_w$ can be seen as ${\mathcal{P}}_w$ without the blocked paths and adding the augmented paths. Usually, augmented paths are longer and less preferred by passengers. For a specific OD pair $w \in \mathcal{W}_I$, the total path throughput of $w$ after the incident, denoted as $\Tilde{T}_w$, should be less than or equal to that before. Therefore, we define $\Tilde{T}_w$ as:
\begin{align}
    \tilde{T}_w = \min\{ \sum_{p\in {\Tilde{\mathcal{P}}_w}} A_p ,\; T_w \}  \quad \quad \forall w \in \mathcal{W}_I
\label{eq_Tw_tilde}
\end{align}
This corresponds to our assumption that the throughput during the incident cannot exceed that before the incident for a specific OD pair. Hence, the NRUI for incident $I$ is formulated as:
\begin{align}
    R_{I} = \frac{\sum_{w \in \mathcal{W}_I} \Tilde{T}_w }{\sum_{w \in \mathcal{W}_I}{T}_w}
\label{eq_RI}
\end{align}
where the numerator (denominator) is the total throughput of available paths after (before) the incident. Since $T_w \geq \Tilde{T}_w$ for all $w \in \mathcal{W}_I$ by definition, we have $0\leq R_{I} \leq 1$. $R_{I} = 1$ means the capacities before and after the incident are the same, suggesting that the incident does not deteriorate the function of the network (i.e. the network is fully redundant under incident $I$). $R_{I} = 0$ means no alternative paths are available during the incident (i.e. the network has no redundancy under incident $I$) 

\begin{figure}[H]
\centering
\includegraphics[width=0.7\linewidth]{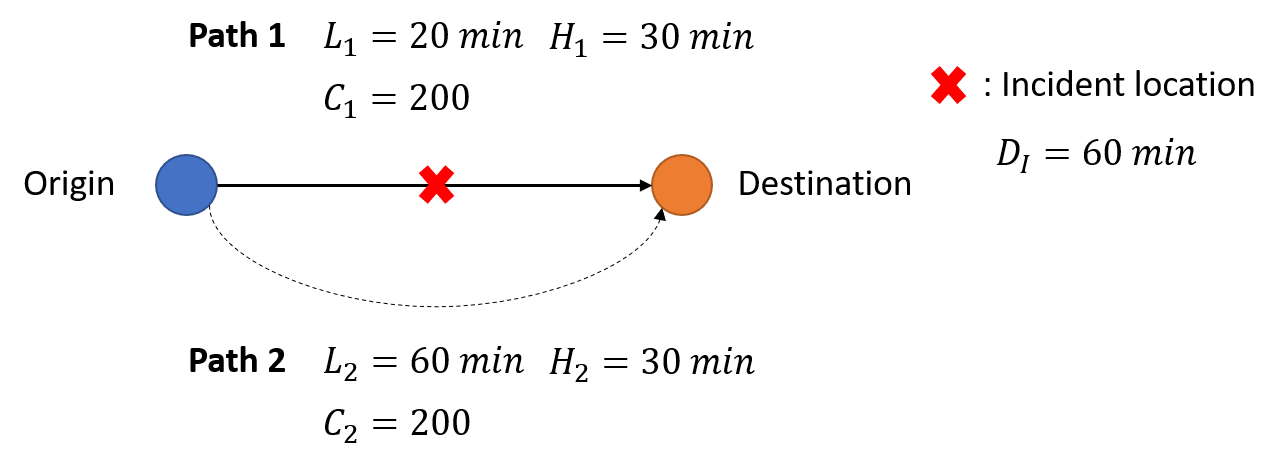}
\caption{Example of network redundancy calculation}
\label{fig_RI_example}
\end{figure}

For better understanding of the index, we present a small numerical example to show how $R_I$ is calculated. As shown in Figure \ref{fig_RI_example}, consider a system with only one OD pair $w$. Path 1 and 2 are primary and alternative paths, respectively, where $\mathcal{P}_w = \{1\}$ and $\Tilde{\mathcal{P}}_w = \{2\}$ (i.e. before the incident path 2 is not chosen by passengers). The attributes of two paths are shown in the figure. For path 1, there are $\floor{D_I/H_p} = 2$ vehicles dispatched. The first vehicle has traveled for $\min\{D_I,L_1\} = 20$ minutes and reached the destination. The second vehicle, which was dispatched 30 minutes later, can also travel for $\min\{D_I - H_1,L_1\} = 20$ minutes and reach the destination. Therefore, these two vehicles successfully carried $400$ passengers to the destination. According to Eq. \ref{eq_Ap}, we have 
\begin{align}
A_1 = \frac{\min\{D_I,L_1\}}{L_1D_I}C_1 + \frac{\min\{D_I - H_1,L_1\}}{L_1D_I}C_1 = 400 \text{ passengers/hour}
\label{Eq_A_1}
\end{align}

In terms of path 2, similarly, there are two vehicles dispatched during the incident. The first vehicle can reach the destination and the second one can only finish $\frac{30}{60}$ of its journey. Therefore, 
\begin{align}
A_2 = \frac{\min\{D_I,L_2\}}{L_2D_I}C_2 + \frac{\min\{D_I - H_2,L_2\}}{L_2D_I}C_2 = 300 \text{ passengers/hour}
\label{Eq_A_2}
\end{align}
The two terms in Eq. \ref{Eq_A_2} represent the number of passengers carried (successfully and partially) by the two vehicles in path 2 per time unit. 

The redundancy index for this single network under incident $I$ is 
\begin{align}
    R_{I} = \frac{\tilde{T}_w}{T_w} = \frac{A_2}{A_1} = \frac{300}{400} = 0.75
\label{eq_RI_test}
\end{align}
which means during the incident when passengers start to use path 2, the system maintains 75\% of its original capacity. For comparison purpose, if one follows the typical definition of path capacity and calculate $A_p$ as $\frac{C_p}{H_p}$, the redundancy index will be 1 because $C_1 = C_2$ and $H_1 = H_2$ in this example, which is obviously unreasonable because it implies that the system maintains 100\% capacity and the incident has no impact.  

It is worth noting that we illustrate the definition of the NRUI assuming a fixed vehicle capacity $C_p$. Actually, $C_p$ can be defined as the ``available capacity'' in a vehicle considering the onboard passengers. In this way, the NRUI can also capture the \textbf{impact of demand}. The ``available capacity'' can be calculated for the specific day of the incident using a transit assignment model \citep{mo2020capacity}. Alternatively, the average ``available capacity'' can also be used. In the case study, due to lack of data, the total capacity of a vehicle is used. 

% ()
\subsubsection{Headway analysis}

Headways are important indicators of the level of service for transit systems. Analyzing headway patterns during an incident can provide direct information about how services are reduced by the incident. As mentioned in Section \ref{intro}, AVL data provide the headway of each station in the urban rail system. In this study, we calculate the headway temporal distribution for lines of interest to capture the impact of incidents. 

Let us divide the analysis time period into several intervals with equal length. Denote the headway on station $i$ of trip $j$ as $H_{i,j}$ (i.e. the length of interval between trip $j$ and $j-1$ and $H_{i,1} = 0$). Suppose line $l$ has two directions, inbound and outbound. The headway of line $l$ \emph{outbound} at time interval $\tau$ is calculated as
\begin{align}
    H_{l,\tau}^{\text{out}} = \frac{\sum_{i \in \mathcal{S}_{l}^{\text{out}}}\sum_{j \in \mathcal{R}_{i, \tau}  } H_{i,j}}{\sum_{i \in \mathcal{S}_{l}^{\text{out}}} |\mathcal{R}_{i, \tau}| -1}
\label{eq_hw_line}
\end{align}
where $\mathcal{S}_{l}^{\text{out}}$ is set of outbound stations in line $l$. $\mathcal{R}_{i, \tau}$ is the set of trips passing through station $i$ during the time interval $\tau$. Eq. \ref{eq_hw_line} implies the headway of a line is calculated as the mean of all stations along the line. The inbound headway, $H_{l,\tau}^{\text{in}}$, is calculated in a similar way by replacing $\mathcal{S}_{l}^{\text{out}}$ with $\mathcal{S}_{l}^{\text{in}}$. The headway distributions of both normal days and the incident day are calculated for comparison. 

\subsection{Demand analysis}

\subsubsection{Passenger flow analysis}
AFC data record passengers' tap-in information in bus and rail systems (tap-out is not available in this study). These transactions can capture passengers' route choices during an incident if they use the transit system again  \citep{mo2020assignment}. Therefore, analyzing AFC data can help understand passenger flow redistribution during an incident. 

At the station level, we calculate the number of tap-in passengers at the stations in the incident area, and compare the values on the incident day and normal days. The difference in this number is an indicator of the impact of the incident on passenger flow redistribution. Stations with high demand increase reflect passengers' choices after the incident. Similarly, at the line level, we calculate the number of tap-in passengers for lines near the incident area for both the incident day and normal days. Line-level demands are calculated as the sum of all station-level demands in corresponding lines.

Note that we assume the number of tap-in passengers is approximately normally distributed. Hence, if the incident day demand is beyond the $\pm 2\times$standard deviation of the normal day demand, we say that a significant difference is observed (i.e., the impact of the incident is significant).

\subsubsection{Individual behavior analysis}\label{sec_method_indi}
Passengers may make different mode choices after the incident. One important question is how the characteristics of the passengers influence their mode choices. This is typically using data from surveys. In this study, we propose a method based on conveniently available AFC data for individual behavior analysis.

The proposed approach consists of two steps: a) inferring individual's mode choice and b) extracting samples' characteristics. We infer individual choices using AFC data. In this study, only two choices are considered: 1) using transit and 2) other (including canceling trips and using other travel modes). This is because these two options can be confidently identified using AFC data and they are important for transit operators. Since passenger travel patterns in transit systems show high irregularity \citep{goulet2017measuring}, it is more convenient to identify the behavioral changes of \emph{regular passengers} \citep{mojica2008examining}. In this study, we define regular passengers as those who use the public transit system every normal day and have the same travel trajectories. Note that, as normal days have the same day of week as the incident day, regular passengers are not necessarily frequent users as they may only use the system on a specific day of week. For example, if an incident happened on Friday, a passenger who only uses the public transit system on Friday (i.e. on each normal day) is a regular passenger. But since he/she only uses the transit once a week, he/she may not be a frequent user. Mathematically, let us denote the trajectories of passenger $i$ in normal day $k$ as $\mathcal{T}^{i,k} = \{(o_1^{i,k},d_1^{i,k},t_1^{i,k}),...,(o_{N^{i,k}}^{i,k},d_{N^{i,k}}^{i,k},t_{N^{i,k}}^{i,k})\}$, where $o_n^{i,k}$, $d_n^{i,k}$, and $t_n^{i,k}$ are the origin, destination and start time of the $n$-th trip, respectively ($t_1^{i,k}<...<t_{N^{i,k}}^{i,k}$). $N^{i,k}$ is the total number of trips in normal day $k$ for passenger $i$. The set of regular passengers is defined as $\{i \in \mathcal{I}\;|\; o_n^{i,k} = o_n^{i,k'}, d_n^{i,k} = d_n^{i,k'}, t_n^{i,k} \in [\bar{t}_n^{i} - \sigma_n^{i},  \bar{t}_n^{i} + \sigma_n^{i}], N^{i,k} = N^{i,k'}, \forall k,k'\in \mathcal{K}, k \neq k'\}$, where $\mathcal{I}$ is the set of all passengers, $\bar{t}_n^{i}$ and $\sigma_n^{i}$ are the mean and standard deviation of the start time of trip $n$ for passenger $i$ over all normal days. $\mathcal{K}$ is the set of all normal days considered in this study. This means regular passengers have the same number of trips and corresponding origins and destinations in each normal day (for tap-in only AFC systems, destinations are not considered). And the corresponding trip start times in each normal day are stable (i.e. within a standard deviation). Hence, if these passengers had different travel patterns on the incident day, most likely they would be affected by the incident and chose a new travel mode. Denote the trip sequence of passenger $i$ on the incident day as $\mathcal{T}^{i,\text{In}} = \{(o_1^{i,\text{In}},d_1^{i,\text{In}},t_1^{i,\text{In}}),...,(o_{N^{i,\text{In}}}^{i,\text{In}},d_{N^{i,\text{In}}}^{i,\text{In}},t_{N^{i,\text{In}}}^{i,\text{In}})\}$. And let $[T_e, T_s]$ be the incident period, where $T_e$ and $T_s$ is the incident start and end time. The mode choice during the incident for a regular passenger $i$ is denoted as $Y_i$. We infer $Y_i$ as follows:
\begin{itemize}
\item $Y_{i}=\text{``Transit''}$ if 1) there are additional transit trips (compared to that in a normal day) during the incident period or 2) there are changes of tap-in stations during the incident period. The first condition implies that the regular passenger may have transferred to a nearby rail station or bus stop, with more transit trips than usual. The second condition implies that the regular passenger may have changed to a different rail line or bus route in response to the incident.  Let $\mathcal{T}_I^{i,k} = \{(o_n^{i,k},d_n^{i,k},t_n^{i,k})\in\mathcal{T}^{i,k}\;|\; T_s \leq \bar{t}_n^{i} \leq T_e  \}$ and $\mathcal{T}_I^{i,\text{In}} = \{(o_n^{i,\text{In}},d_n^{i,\text{In}},t_n^{i,\text{In}})\in\mathcal{T}^{i,\text{In}}\;|\; T_s \leq t_n^{i,\text{In}} \leq T_e  \}$ be sub-sequences of trips within the incident period (i.e. $[T_s, T_e]$) on a normal and the incident day, respectively. Mathematically, the first condition can be expressed as: $|\mathcal{T}_I^{i,\text{In}}| > |\mathcal{T}_I^{i,k}|$ and the second: $\exists n$ s.t. $o_n^{i,k}\neq o_n^{i,\text{In}}$, where $(o_n^{i,k},d_n^{i,k},t_n^{i,k}) \in \mathcal{T}_I^{i,k}$ and $(o_n^{i,\text{In}},d_n^{i,\text{In}},t_n^{i,\text{In}}) \in \mathcal{T}_I^{i,\text{In}}$. Note that $k\in\mathcal{K}$ can be any normal day because the trajectories for all normal days are the same for a regular passenger by definition. 
\item $Y_{i}=\text{``Other''}$ if the transit trips that are supposed to happen during the incident period on the normal days disappear on the incident day. This means that the regular passengers may change to other modes or cancel their trips. Mathematically, this can be expressed as  $|\mathcal{T}_I^{i,\text{In}}| < |\mathcal{T}_I^{i,k}|$.
\end{itemize}
Other regular passengers without the above behavior may not be affected by the incident or have other choices that are hard to be identified (e.g., transfer to another line without leaving the system), which are not considered in the analysis.

In the second step, the characteristics of each regular passenger (i.e. demographics and trip information) are extracted. We aim to use information that is available in AFC and sale transaction data as a proxy to passengers' socio-demographics. 

Since regular passengers have consistent travel trajectories, we can infer their home locations as the tap-in rail station or bus stop of the first trip on a normal day (i.e. $o_1^{i,k}$ for any $k\in\mathcal{K}$). Given the station/stop location, we can obtain the median household income in passenger $i$'s neighborhood or census tract using census data. Living in a high-income or low-income neighborhood can be a proxy of passengers' income. AFC data can also provide passengers' fare status information, such as whether the passenger is in a reduced fare status. Reduced fare status users are usually students, seniors, and people with disabilities. This information is also a proxy for socio-demographic characteristics.

Sale transaction data provide the historical add-value transactions of passengers. We extract three variables in this study: total added value per year, add-value frequency (i.e. number of add-value transactions per year), and maximum single added value in a year. The first two variables reflect the passenger's dependence on and familiarity with public transit and part of their income information. The last variable can also be used to some extent as a proxy for income because low-income people may not be able to deposit a large amount of money in the smart card at once. We denote all this ``proxy'' demographic information for passenger $i$ as $X_i$.  

The characteristics of passenger $i$'s trip (denoted as $Z_i$) during the incident may also affect mode choices. We define the incident-related trip (trip ID denoted as $n^*$) as the first trip with $\bar{t}^{i,k}$ in the incident period. Mathematically, $n^* = \argmin_n \{n= 1,...,N^{i,k}\;|\;\bar{t}_n^{i} \in [T_s,T_e]\} $. Since regular passengers are supposed to have stable travel patterns, $o_{n^*}^{i,k}$ and $d_{n^*}^{i,k}$ should be the intended origin and destination for passenger $i$ on the incident day. Based on $(o_{n^*}^{i,k}, d_{n^*}^{i,k})$, two trip-related variables are considered. The first is whether the $d_{n^*}^{i,k}$ is downtown, which is a proxy for work trips. Note that for a tap-in only system, $d_{n^*}^{i,k}$ can be inferred from a destination estimation model \citep{barry2002origin, zhao2007estimating, gordon2013automated}. The second variable is \emph{OD-based redundancy}, defined as 
\begin{align}
    R^{\text{OD}}_i = \frac{\Tilde{T}_{w_i}}{{T}_{w_i}}
\end{align}
where $R^{\text{OD}}_i$ is the {OD-based redundancy} for passenger $i$, measuring the availability of alternative transit services for the specific OD pair during an incident. ${w_i} = (o_{n^*}^{i,k}, d_{n^*}^{i,k})$ is passenger $i$'s OD pair for the incident-related trip. It is worth noting that $R^{\text{OD}}_i$ can be seen as the NRUI for the case of a single OD pair. 

In this study, we use a binary logit model \citep{ben1985discrete} to better understand the main factors that impact choice $Y_i$. Let the utility of mode $j$ for passenger $i$ be $U_{ij}$.
\begin{align}
    U_{ij}= \text{ASC}_j + \alpha_j X_i + \beta_j Z_i + \epsilon_j
\end{align}
where $\text{ASC}_j$ is the alternative specific constant (ASC) for mode $j$. $\epsilon_j$ is the error term that is assumed to be Gumbel distributed. $\alpha_j$ and $\beta_j$ are the vectors of parameters to be estimated. The probability of passenger $i$ choosing mode $j$ is
\begin{align}
    \mathbb{P}(Y_i = j) =\frac{\exp(\text{ASC}_j + \alpha_j X_i + \beta_j Z_i)}{\sum_{j' \in \mathcal{C}}\exp(\text{ASC}_j + \alpha_j X_i + \beta_j Z_i)}   \quad \quad \forall j \in \mathcal{C}
\end{align}
where $\mathcal{C} = \{\text{``Transit''}, \text{``Other''}\}$ is the choice set. 

\begin{figure}[H]
\centering
\includegraphics[width=0.8\linewidth]{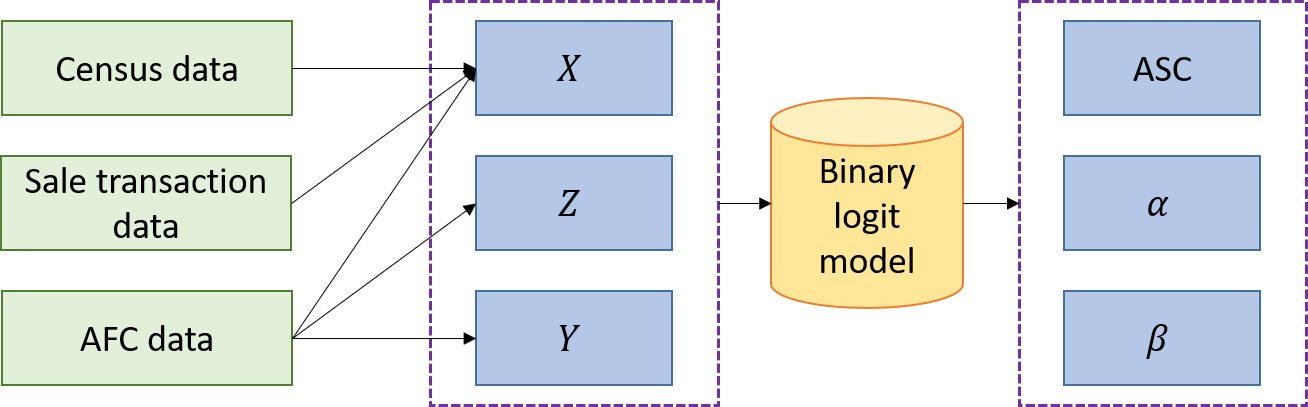}
\caption{Summary of the individual behavioral analysis model}
\label{fig_individual_model}
\end{figure}

The approach of the individual behavioral analysis model is summarized in Figure \ref{fig_individual_model}.  

% The model is implemented by Python \texttt{Biogeme} package \citep{bierlaire2018pandasbiogeme}.

\section{Application}\label{sec_case}

\subsection{Chicago Transit System}
\begin{figure}[H]
\centering
\includegraphics[width=0.7\linewidth]{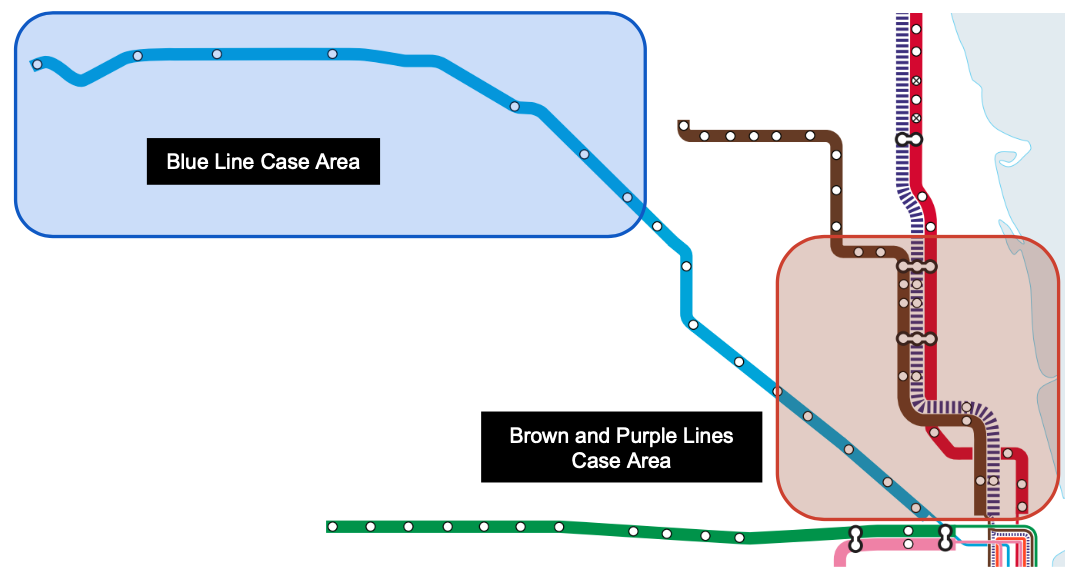}
\caption{CTA rail system map}
\label{fig_cta_rail_map}
\end{figure}

We use incident data from the Chicago Transit Authority (CTA) public transit system for the model application in this section. CTA is the second-largest transit system in the United States, providing services in Chicago, Illinois, and some of its surrounding suburbs. It operates 24 hours each day and is used by 0.84 million bus and 0.81 million train passengers per weekday on average \citep{CTA2019Rider}. The map of the CTA rail system is shown in Figure \ref{fig_cta_rail_map}. The rail system consists of eight lines (named after their color) and the "Loop". The Loop, located in the Chicago downtown area, is a 2.88 km long circuit of elevated rail that forms the hub of the Chicago rail system. Its eight stations account for around 10\% of the weekday boardings of the CTA trains. 

CTA’s AFC system is entry-only, meaning passengers use their farecards only when entering a rail station or boarding a bus, and so no information about a trip’s destination is directly provided. The train tracking system provides train arrival and departure times at each station. 

According to the control center data, CTA experienced a total of 27,198 incidents in 2019. However, around 80\% percent of the incidents have a duration of fewer than 10 minutes. Since small incidents may not affect the system significantly, this study focuses on substantial incidents that lasted longer than 1 hour. Passengers who leave the rail system because of service disruptions need to re-tap in if they decide to use other CTA services (buses or rails). They are only charged a transfer fee. However, no tap-in is needed for shuttle service that may have been deployed in response to the incident. Hence, there is no information for passengers using shuttle buses.

\subsection{Redundancy index}\label{sec_redun}
Prior to analyzing actual incidents, we first present an overview of the CTA system redundancy. As the NRUI is defined based on each incident, for the purpose of this analysis, we assume that a hypothetical incident takes place at a station in the system (one at a time), blocking the track segment that connects the station for 1 hour. Note that if a station has two separate tracks, each track is blocked independently and there will be two incident cases for this station. For example, the Roosevelt station has two different tracks for the Red Line and Purple/Yellow Lines. So two hypothetical incident cases are generated, each corresponding to the interruption of a track. Considering the infrastructure layout of urban rail systems, assuming that incidents occur at the track level is more realistic than simply assuming an incident blocks the whole station as in many previous studies that used graph-based methods \citep{chopra2016network,ye2019assessing}.

\begin{figure}[H]
\centering
\includegraphics[width=0.5\linewidth]{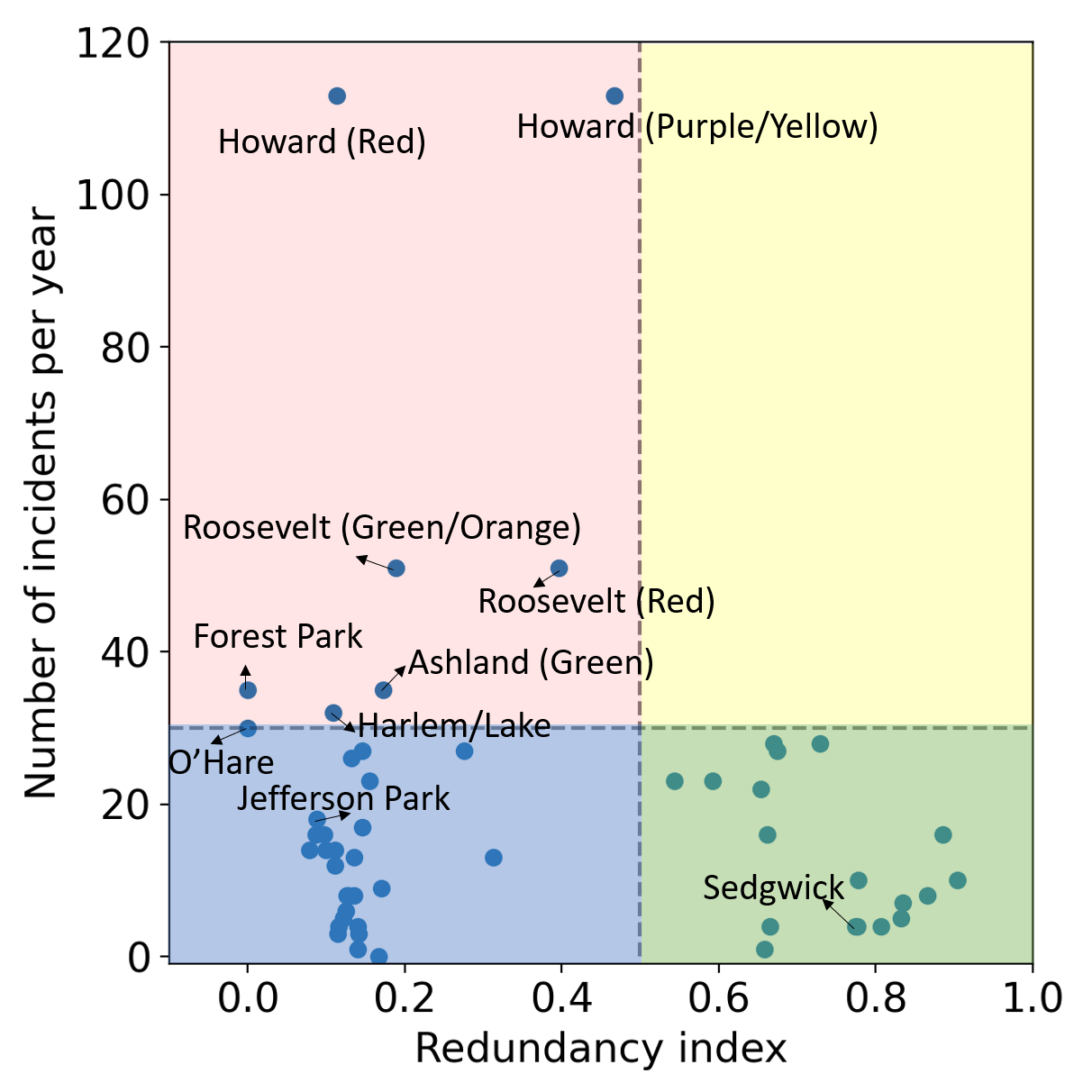}
\caption{Redundancy index v.s. incident occurrence rate. }
\label{fig_cta_redundancy}
\end{figure}

Besides the incident-specific redundancy index, the occurrence frequency of incidents at various stations is also of importance. Figure \ref{fig_cta_redundancy} shows the redundancy index at each station against the number of incidents taking place per year at that station (only incidents with a duration greater than 10 minutes are counted). The combination of the two metrics divides the figure into four sections: 1) Stations in the red section (upper left) have high incident occurrence frequency and low redundancy. These are critical stations in the system where alternative public transit services are limited and service disruptions happen frequently. Transit operators need to prepare strategies in advance for these stations. 2) Stations in the yellow section (upper right) have high incident occurrence frequency and high redundancy. In these stations, passengers are able to seek alternative services during a disruption. Operators need to provide direct information to passengers with suggestions regarding alternatives. 3) Stations in the blue section (lower left) have low incident occurrence frequency and low redundancy. Though incidents may not happen frequently, mitigation plans need to be prepared as there are limited substitutional services. 4) Stations in the green section (lower right) have low incident occurrence frequency and high redundancy. These stations are less critical in terms of incident management compared to stations in other sections. 

Figure \ref{fig_cta_redundancy} shows that most of the stations in the CTA system are in the blue or green sections. And only a limited number of stations are in the red section. This implies that CTA can focus more on some critical stations with adequate incident management strategies. In terms of critical stations (red section), most of them are terminal stations (such as Howard, Forest Park). This is expected as terminal stations usually have more complex infrastructure layouts (i.e. more prone to failures) and are usually located in suburban areas (i.e. fewer alternative services and low redundancy index). Backup shuttle services can be provided in these stations.

\subsection{Rail disruption cases}

Since the location of the incidents may influence their impact, we selected two incidents at locations with high and low redundancy, respectively, for comparative analysis. 

% Since the location of the incidents may influence the impact on the entire system, we selected two case studies of interest with both low and high redundancy to compare. 

% The first case was on the Blue Line in outer Chicago near Jefferson Park, where a 3rd rail was reported to have been dislodged. The second case was close to the city core at Sedgwick Station, on the Brown and Purple Lines, where two trains collided with each other. To ensure that each incident had either low or high redundancy, we calculated the redundancy for each.

% To get this, we looked at all the Origin-Destination (OD) pairs with paths that crossed both blocked links, defined as N1. For each of those OD pairs, we generated an alternative path using rail (excluding bus), defined as N2. The redundancy was then calculated by dividing N2/N1, yielding a value between 0 and 1, with 0 meaning not redundant, and 1 meaningfully redundant.

% The Blue Line Jefferson Park case scored a redundancy of 0, which is understandable because there are no alternative rail connections. The Brown Line Sedgwick Case had a redundancy of 0.87, meaning that 87\% of OD pairs using these block links still have available rail routes.

% We felt these two cases would be most effective in testing and examining both a system-wide and individual-level analysis of the impact of incidents, using AFC data.

\subsubsection{Brown and Purple Lines Sedgwick incident}
\begin{figure}[H]
\centering
\includegraphics[width=2.5 in]{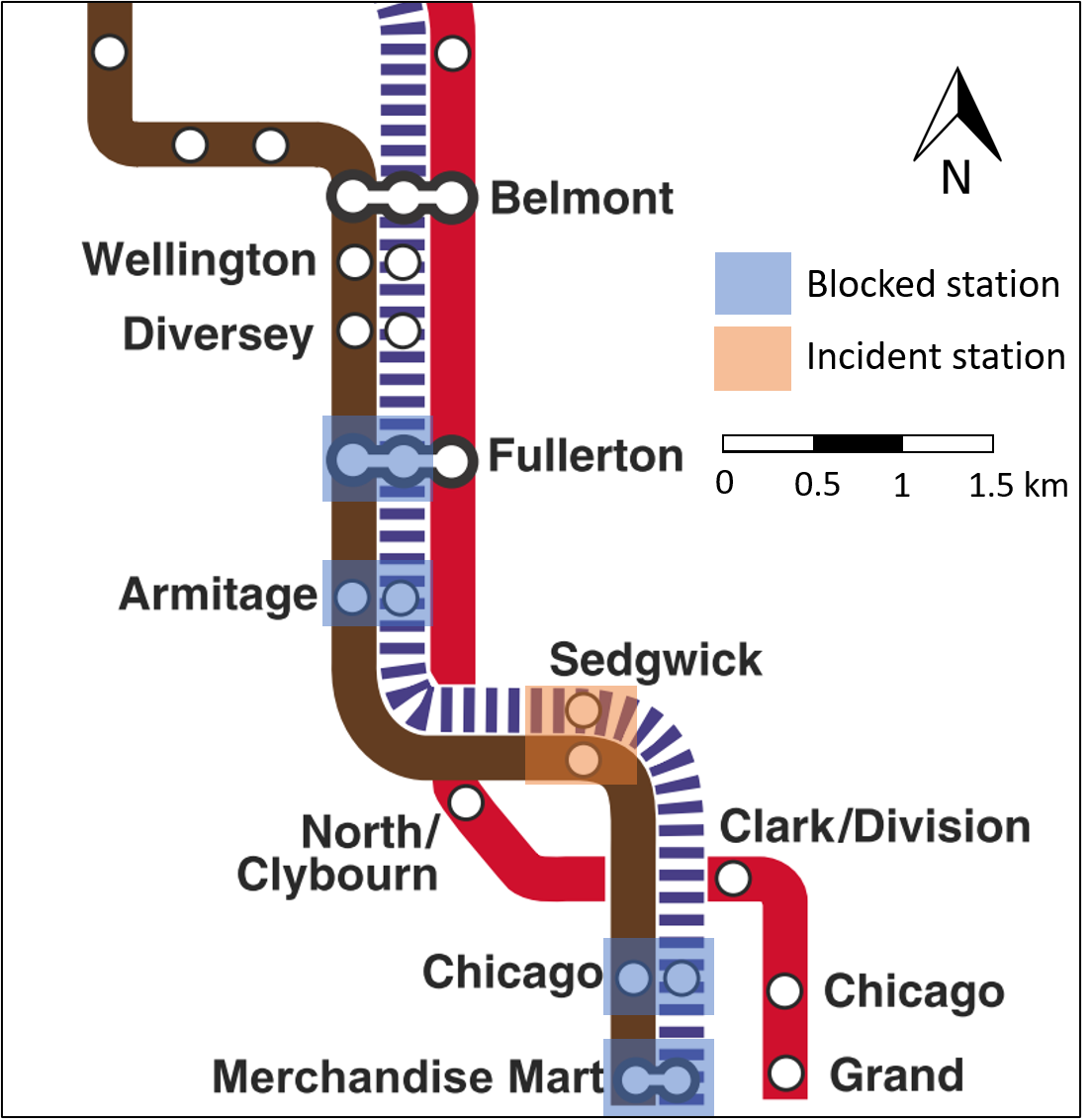}
\caption{Incident diagram of Brown Line Sedgwick case}
\label{fig:brownlinemap}
\end{figure}

On September 24 (Tuesday), 2019, at 9:09 AM, a Purple Line train collided with a Brown Line train at the Sedgwick station. The incident caused a number of stations to be blocked and closed in both Brown and Purple Lines since these two lines share the same track in this area. The impacted stations were Fullerton and Armitage to the north and Chicago and Merchandise Mart (MM) to the south. Southbound trains short turned at Fullerton, while northbound trains short turned at MM. At 9:28 AM, 19 minutes after the incident started, bus substitution service began between Fullerton to MM. Service resumed at all blocked stations at 10:19 AM, 70 minutes after the start of the incident. The incident on the Brown and Purple Lines is a high redundancy case because the Red Line is a good substitution for the incident location (See Figure \ref{fig:brownlinemap}).

\subsubsection{Blue Line Jefferson Park incident}
\begin{figure}[H]
\centering
\includegraphics[width=4 in]{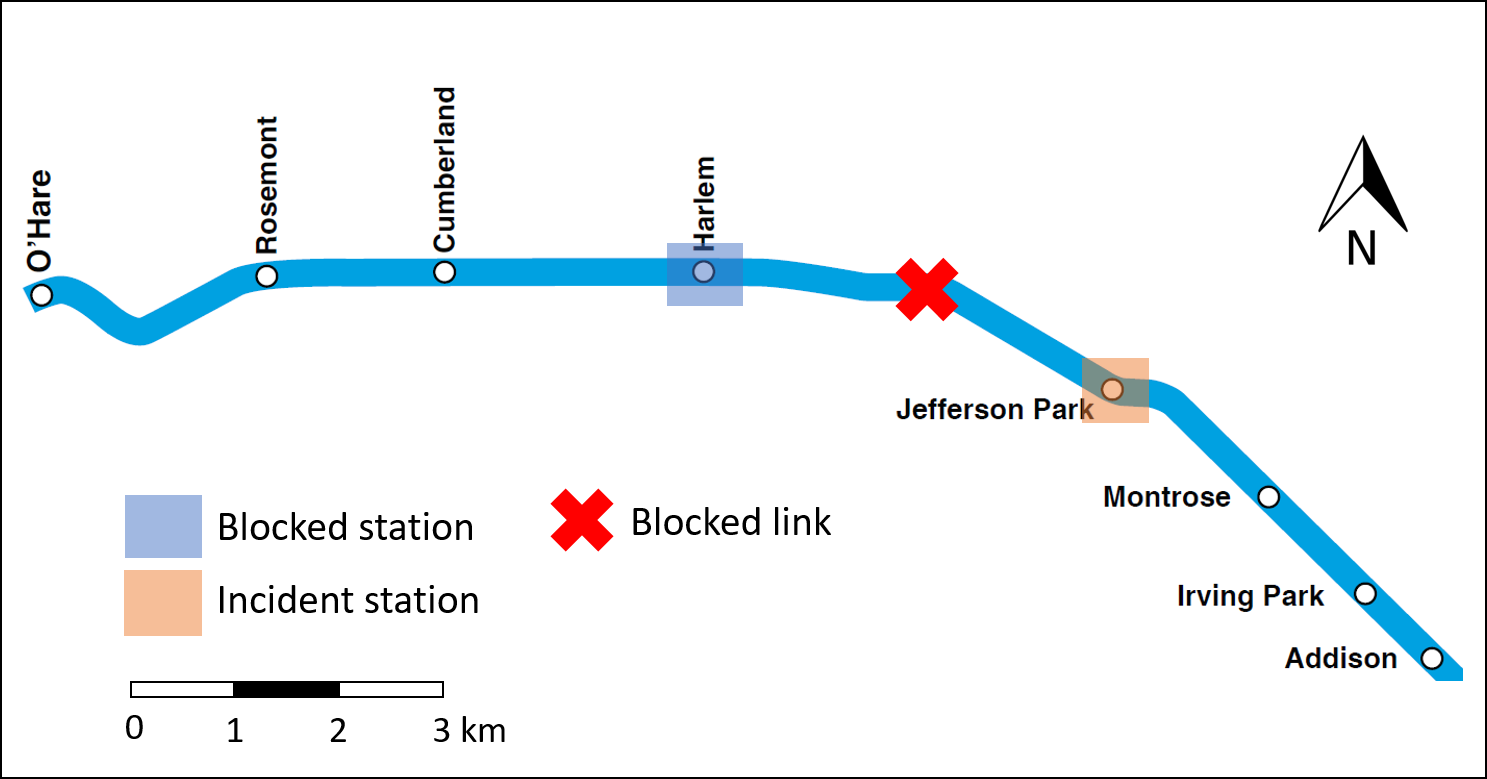}
\caption{Incident diagram of Blue Line Jefferson Park case}
\label{fig:bluelinemap}
\end{figure}

On February 1 (Friday), 2019, at 8:14 AM, the inbound track Blue Line between Harlem and Jefferson Park was closed due to infrastructure problems. All trains in the Blue Line were suspended. CTA used the remaining single-direction track to serve trains from both directions in the incident link. At 9:03 AM, 49 minutes after the incident, single track operations commenced between Harlem and Jefferson Park, with shuttle service starting 7 minutes later. At 9:40 AM, all inbound trains succeeded to move under the single-track operation. At 12:09 PM, the full line was reopened. The entire incident lasted 4 hours and 9 minutes. The incident on the Blue Line is a low redundancy case because the Blue Line is far away from other rail lines with limited alternative services  (see Figure \ref{fig:bluelinemap}).

% Passengers had the option of exiting the system and taking a non-public transit mode, like a TNC (like Uber or Lyft), walk, or bike. They could also take the bus replacement service between the two stations. They could remain on the train and wait for the train's turn to traverse the single track area. Lastly, riders could switch to a nearby bus line and complete their journey on that. 

\section{Analysis}\label{sec_result}

The framework discussed in Section \ref{sec_method} was used for the analysis of the cases. For each case, the results are organized from supply to demand analysis. The individual choice analysis is conducted based on samples of affected passengers from two incidents.

\subsection{Brown and Purple Line incident analysis}
\subsubsection{Redundancy index}
The NURI (Eq. \ref{eq_RI}) for the Brown and Purple Line case is 0.732, meaning that the transit system maintains 73.2\% transporting capacity for the Brown and Purple Lines incident during the incident period. The high redundancy of the Brown and Purple Lines incident is as expected. In the incident area, the Red Line is almost parallel with the Brown and Purple Lines. In addition, there exist many south-bound bus routes going to Downtown Chicago (see Figure \ref{fig_brownline_demand_station}). This implies that during the Brown and Purple Line incident, CTA can focus on guiding passengers to find alternative services. Some information dissemination strategies need to be applied, such as route and transfer recommendations.

\subsubsection{Headway analysis}\label{sec_brown_headway}
\begin{figure}[H]
\centering
\subfloat[Brown Line (southbound)]{\includegraphics[width=0.33\textwidth]{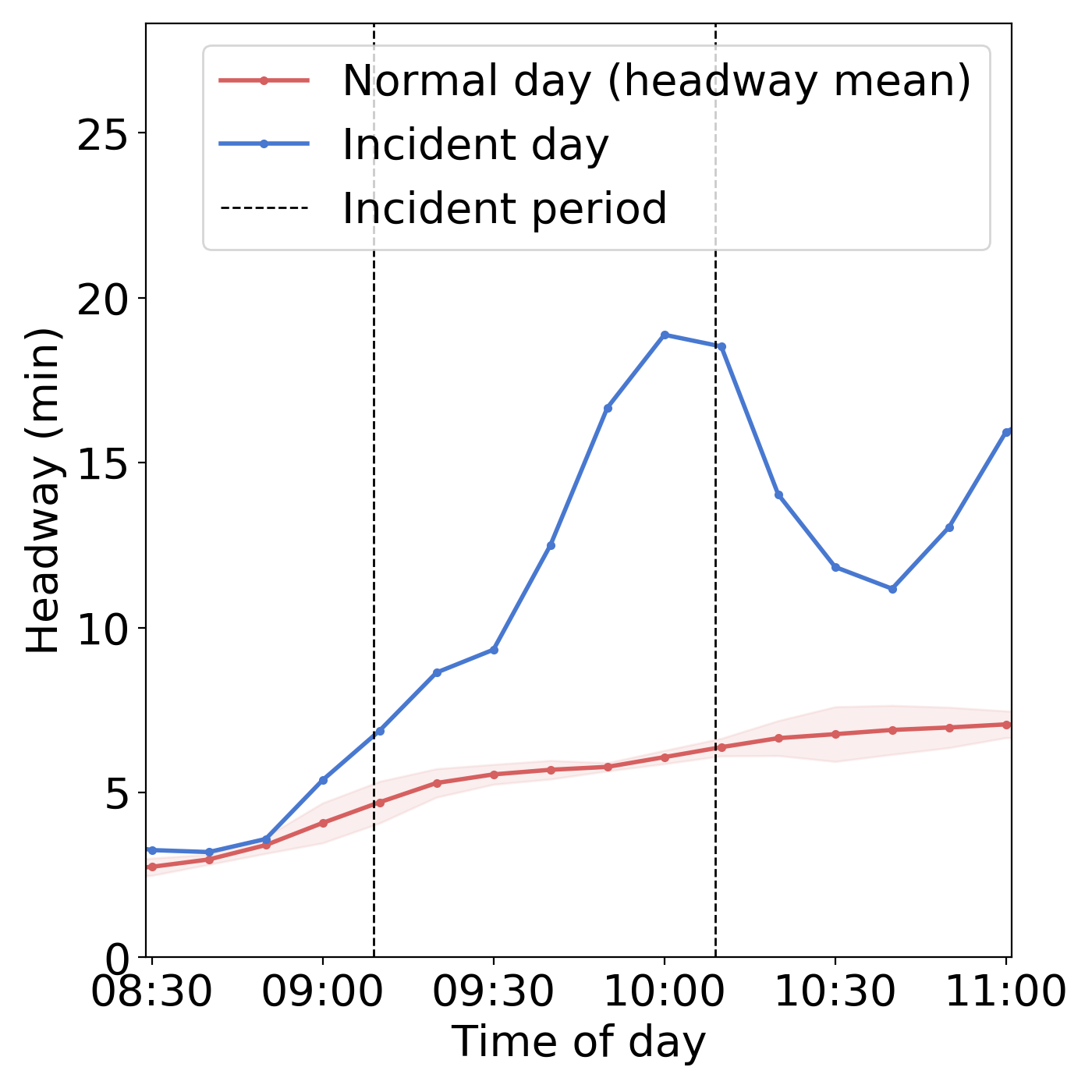}\label{fig_brownsupply_1}}
\hfil
\subfloat[Purple Line (southbound)]{\includegraphics[width=0.33\textwidth]{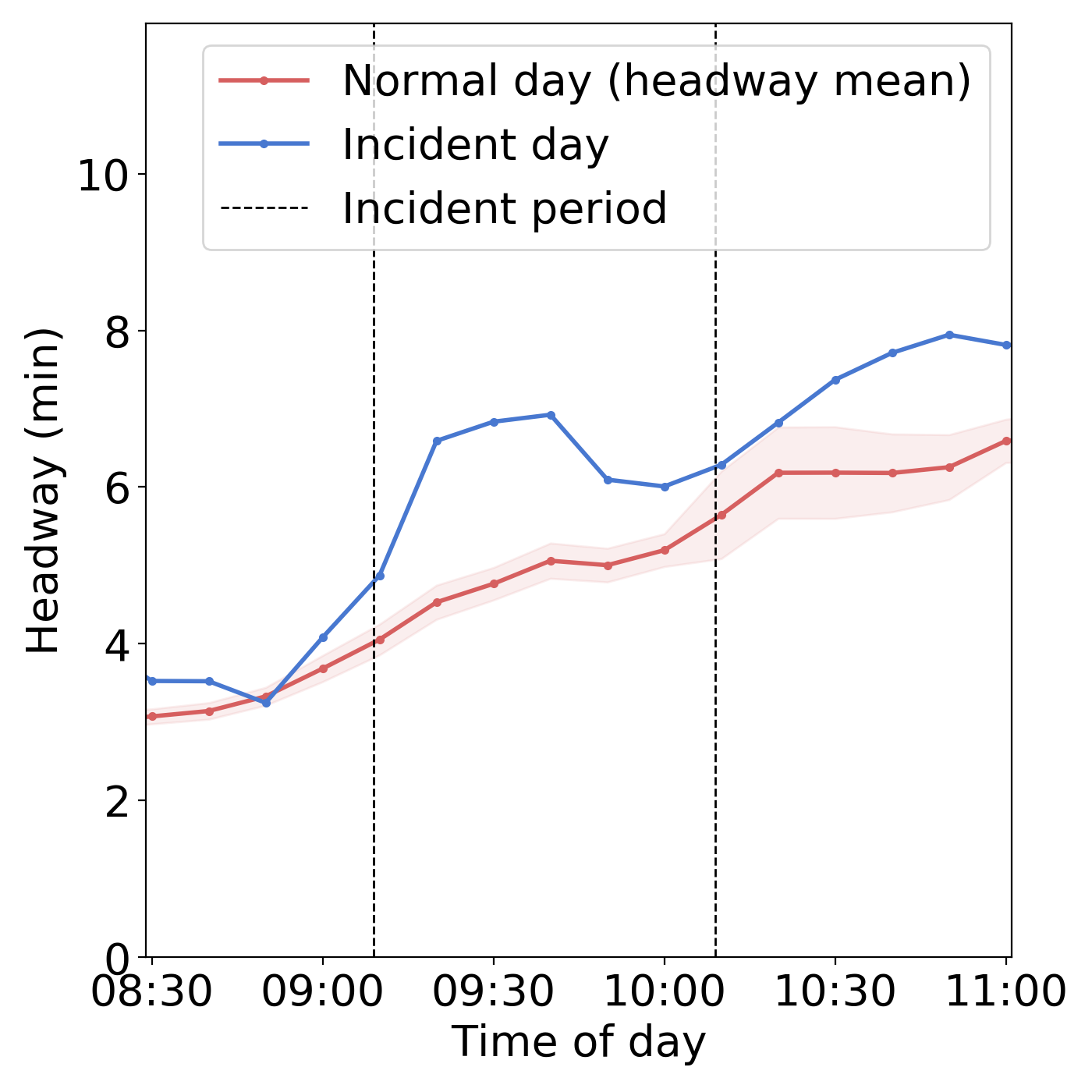}\label{fig_brownsupply_2}}
\hfil
\subfloat[Red Line (southbound)]{\includegraphics[width=0.33\textwidth]{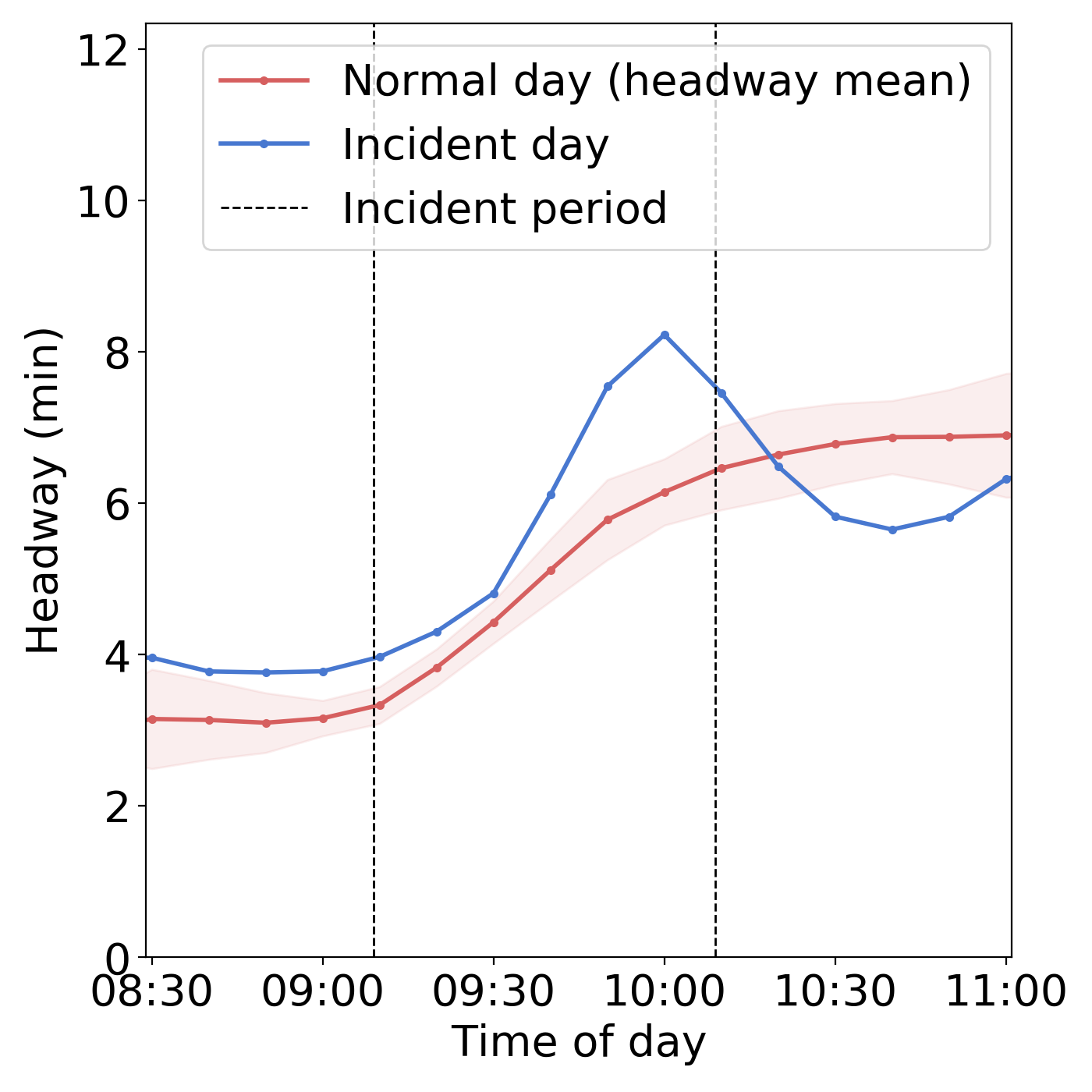}\label{fig_brownsupply_3}}
\caption{Headway temporal distribution (Brown and Purple Lines Sedgwick incident). The shade around normal day lines indicates $\pm$standard deviation (same for all following figures)}
\label{fig:supply_brownline}
\end{figure}

The headway analysis results from the Brown and Purple Line Sedgwick incident are summarized in Figure \ref{fig:supply_brownline}. The shade around normal day lines indicates $\pm$standard deviation (same for all the following figures with shades around normal day lines).  The line-level headway is calculated as Eq. \ref{eq_hw_line}. We selected three lines with directions of interest to analyze. Recall that the Brown and Purple Lines share tracks in the incident area, while the Red Line runs on separate tracks in the incident area but shares tracks further north of the line. 

A rise in southbound headways for both the Brown and Purple Lines are observed (Figures \ref{fig_brownsupply_1} and \ref{fig_brownsupply_2}) and the changes are significant (i.e., beyond the two standard deviation range). This is as expected because these two lines are blocked due to the incident. On average, headway increases from 5 minutes to 15 minutes in Brown Line and from 5 minutes to 7 minutes in Purple Line, implying a reduction of service frequency by 66.7\% and 28.6\% for Brown and Purple Lines, respectively. The Brown Line experiences a continuous increase in headways towards the end of the incident. And we see a decrease in headways once the incident clears. The Purple Line, which has most of its local stops farther away from the incident area, has less disrupted service at the line level, despite sharing tracks with the Brown Line. So its headways deviated less from the normal-day average. 

As shown in Figure \ref{fig_brownsupply_3}, the Red Line experiences little deviation from its normal day service for the first half of the incident (before 9:30 AM), largely because it does not share tracks at the incident location and could run largely uninterrupted. However, halfway through the incident, there is a headway increase spike. This could be caused by two possible reasons: 1) Because of the bad service on Brown and Purple Lines, passengers chose to take the Red Line southbound instead, leading to more passengers and thus the delays at the stations when loading and unloading passengers. 2) The unusual operation (e.g., short-turn) of the trains on Brown and Purple Lines may occupy facilities in the Red Line, resulting in congestion and longer headways. The headway increase in nearby lines implies that the transit operator should pay attention to both incident lines and nearby lines to better serve passengers.

\subsubsection{Passenger flow analysis}\label{sec_brown_flow}
\begin{figure}[H]
\centering
\subfloat{\includegraphics[width=0.5\textwidth]{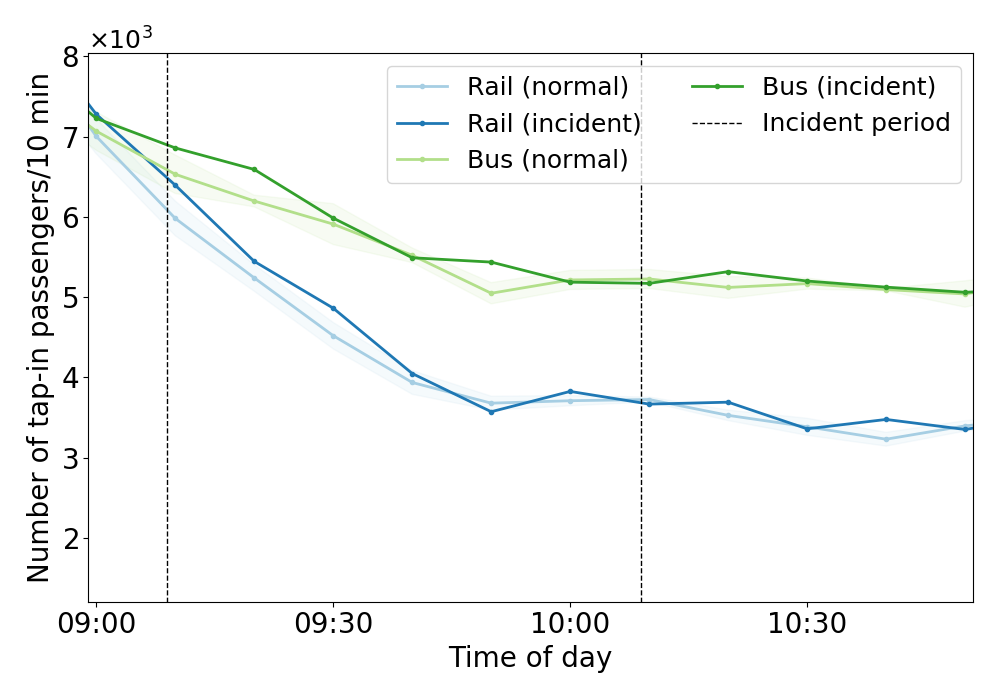}}
\caption{System level passenger flow analysis (Brown and Purple Lines incident).}
\label{fig:systemlevelbrown}
\end{figure}

Passenger flows can be examined at multiple levels, including system-wide, line level, and station level. Figure \ref{fig:systemlevelbrown} shows the total number of tap-in passengers for the bus and rail systems during the Brown and Purple Lines incident. The results show that there is no significant difference between the incident day and normal days for both bus and rail because the demand lines on the incident day are within the $\pm$2 standard deviation range. This implies that though the incident lasted for more than 1 hour and blocked several stations, the impact on the whole system demand is still negligible (i.e., as influential as the inherent demand variations).

\begin{figure}[H]
\centering
\subfloat[Brown Line (blocked)]{\includegraphics[width=0.33\textwidth]{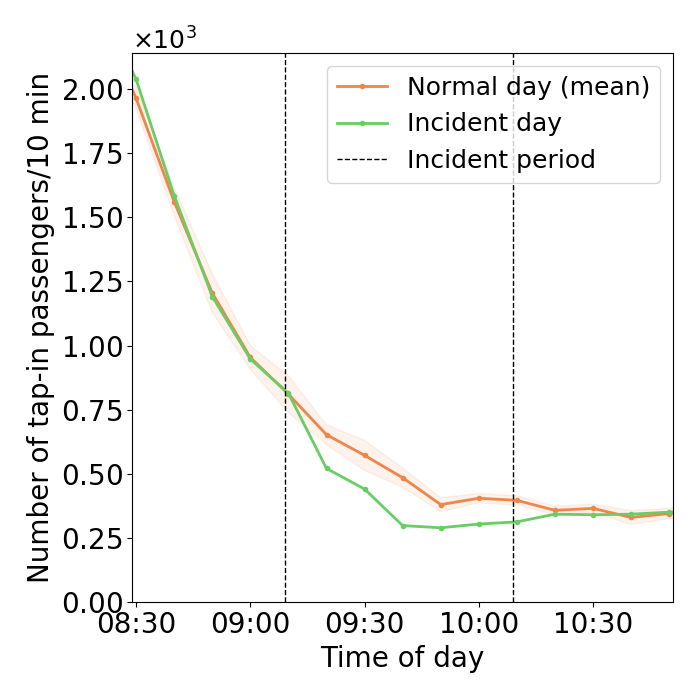}\label{fig_demand_1}}
\hfil
\subfloat[Purple Line (blocked)]{\includegraphics[width=0.33\textwidth]{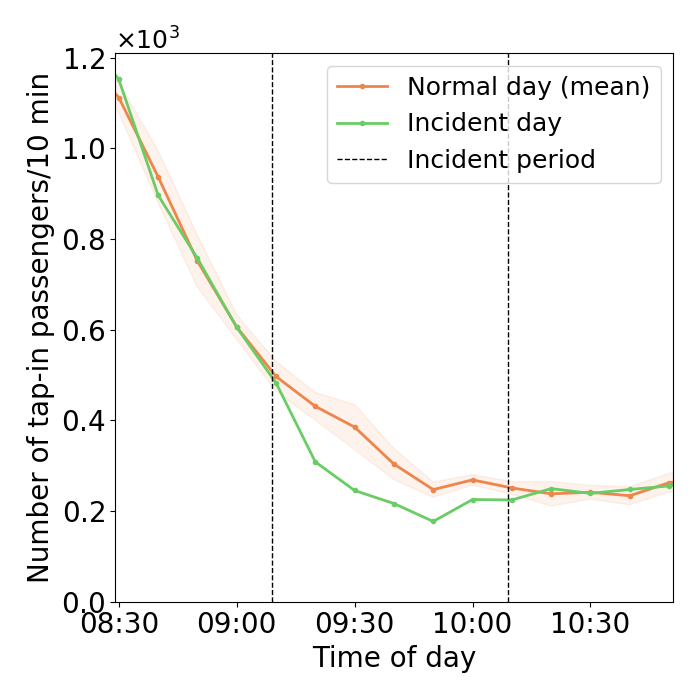}\label{fig_demand_2}}
\hfil
\subfloat[Red Line (open)]{\includegraphics[width=0.33\textwidth]{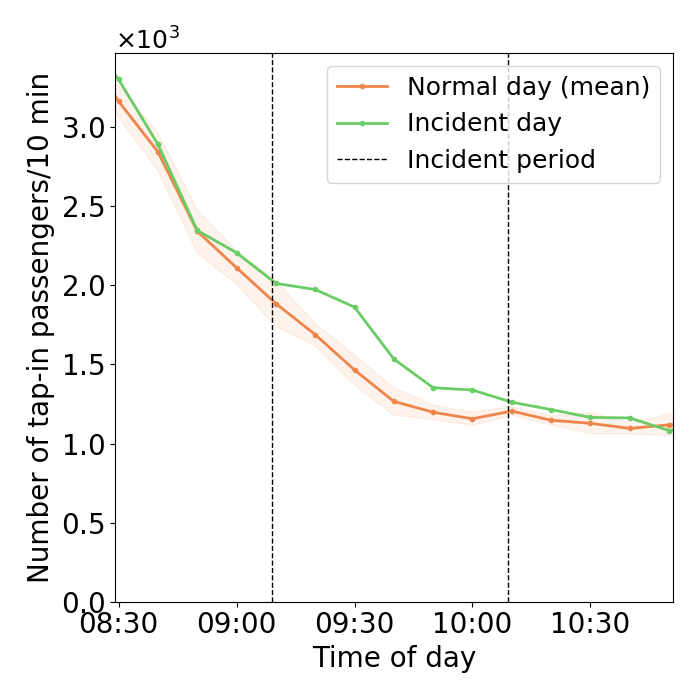}\label{fig_demand_3}}
\caption{Line level passenger flow analysis (Brown and Purple Lines incident)}
\label{fig_linelevelbrown}
\end{figure}

The line-level demand changes for the Brown and Purple Line incident are shown in Figure \ref{fig_linelevelbrown}. As expected, demand on the Brown and Purple Lines (interrupted by the incident) both decreased during the incident and returned to normal after the incident. And the decrease is significant. As the Red Line runs adjacent to the Brown and Purple Lines for a significant portion and is not suspended, we see a significant increase in demand during the incident period with a return to normal after the incident is over.

\begin{figure}[H]
\centering
{\includegraphics[width=0.6\textwidth]{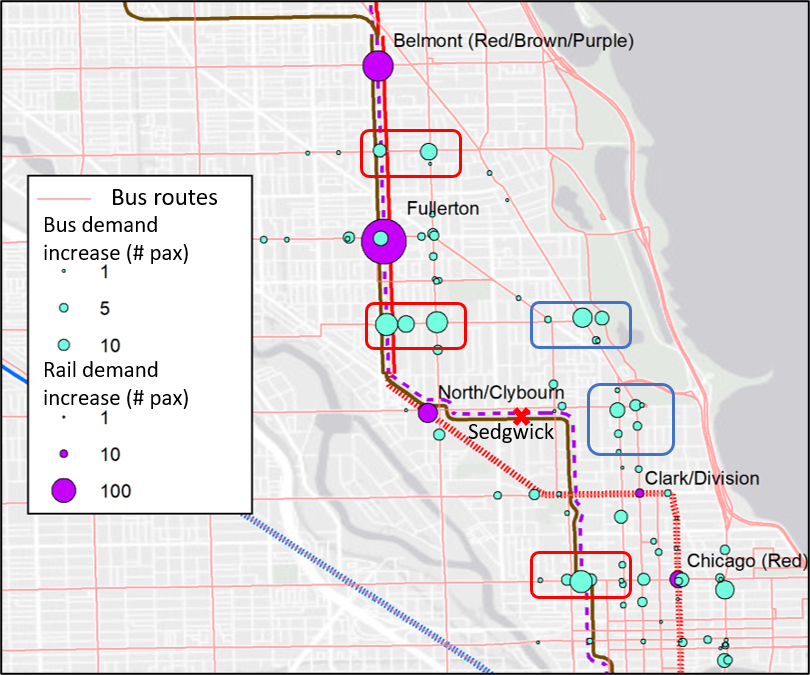}\label{fig_flow_sedgwick_1}}
\caption{Station demand increase patterns (Brown and Purple Lines incident)}
\label{fig_brownline_demand_station}
\end{figure}

We further examine the demand changes at rail and bus stations close to the incident area (shown in Figure \ref{fig_brownline_demand_station}). During the incident, we see an increase in rail demand at Fullerton and Belmont stations that have direct connections to the uninterrupted Red Line. We also see clusters of increased bus demand near the incident lines. Of note are the clusters outlined in red and blue squares. The red clusters represent increased bus demand proximal to blocked stations. These passengers may have transferred directly to nearby bus stops from the blocked line. Additionally, the blue clusters represent increases in bus demand for routes that connect directly to downtown. The increase may be attributed to passengers who live in nearby neighborhoods and change to buses during the incident.

\begin{figure}[H]
\centering
\subfloat[Sedgwick (incident station, blocked)]{\includegraphics[width=0.33\textwidth]{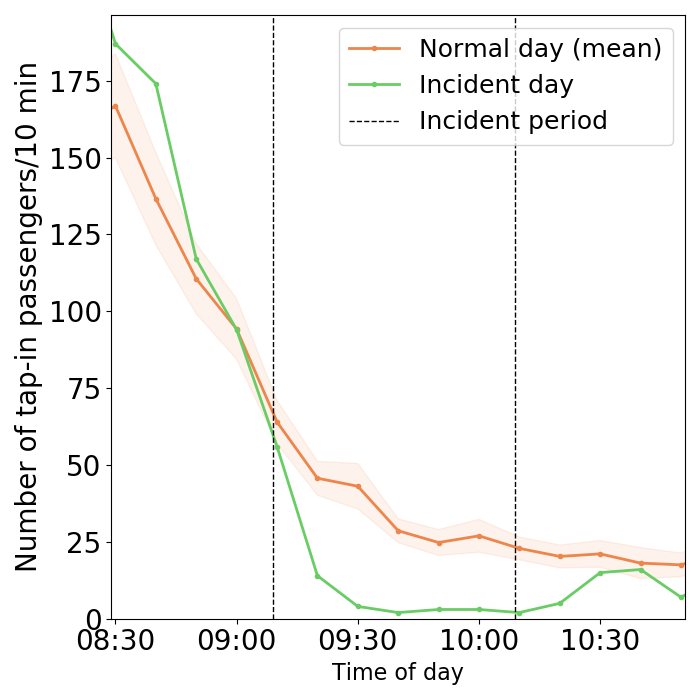}\label{browndemandsedgwick}}
\hfil
\subfloat[Fullerton (partially blocked)]{\includegraphics[width=0.33\textwidth]{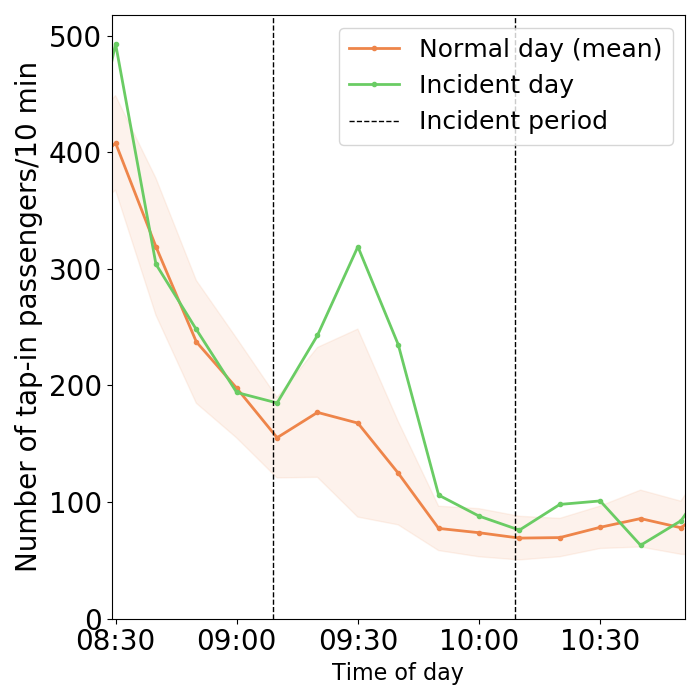}\label{browndemandfullerton}}
\hfil
\subfloat[North/Clybourn (Red Line, open)]{\includegraphics[width=0.33\textwidth]{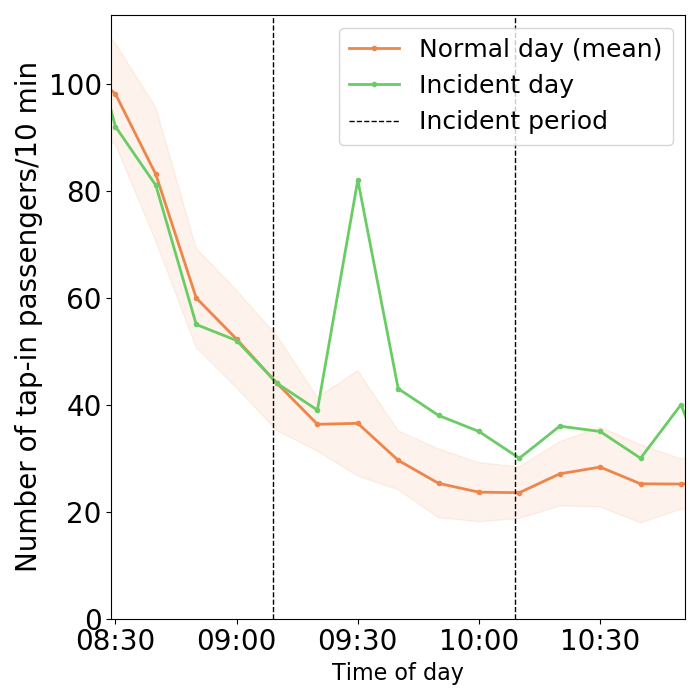}\label{browndemandclybourn}}
\caption{Station level passenger flow analysis (Brown and Purple Lines incident)}
\label{fig:brownlinestationanalysis}
\end{figure}

The total decrease in the number of tap-in passengers in the Brown and Purple Lines is 1,141, while the increase in nearby bus stations and the Red Line is 696 and 1,414, respectively. The demand decrease in the Brown and Purple lines is smaller than the corresponding increase in the Red Line and bus stations. This is probably because some passengers may first tap in the Brown and Purple Lines and then leave (this phenomenon will be illustrated in Figure \ref{fig:brownlinestationanalysis}), which leads to the underestimation of demand decrease in Brown and Purple Lines. For all the 2,110 observed passengers using the alternative services, around one-third of them (696) transfer to buses and two-thirds (1,414) to the Red Line. Note that there may also be many passengers with direct transfers without leaving the system, which cannot be observed from the AFC data.  

Additionally, Figure \ref{fig:brownlinestationanalysis} shows the temporal demand distribution at three stations: Sedgwick (the incident station), Fullerton (a nearby partially blocked station), and North/Clybourn (a nearby station in Red Line that is open). The illustrated trends align with the incident pattern. At Sedgwick (Figure \ref{browndemandsedgwick}), the center of the incident, we see a drastic decrease in demand once the incident starts. As some passengers may not be aware of the incident and accidentally tapped into the station, the demand is not zero during the incident period. After the incident is over, we see a quick recovery in demand. In terms of the Fullerton (Figure \ref{browndemandfullerton}) station, despite it being partially blocked (the tracks of Brown/Purple Lines are blocked but the track of Red Line are not), we see an immediate rise in demand. This indicates that passengers used Fullerton station for the Red Line. Lastly, we see a sharp increase in the number of tap-in passengers at the North/Clybourn station in the Red Line (Figure \ref{browndemandclybourn}), which is within walking distance from Sedgwick station. This implies that passengers from the Brown and Purple Lines may also walk to the Red Line to finish their journey. Additionally, this station gives passengers access to Fullerton station, where they can switch to Brown or Purple Lines trains going northbound. The sharp increase may represent the first wave of transfer passengers. 

The demand analysis is helpful for transit operators to identify passengers' choices, supplement transit on other lines, and inform passengers of better alternatives.

%===========BLUE BELOW, BROWN ABOVE=============

\subsection{Blue Line incident analysis}

\subsubsection{Redundancy index}
The NURI (Eq. \ref{eq_RI}) for the Blue Line Jefferson Park case is 0.093, meaning that the transit system maintains 9.3\% transporting capacity during the incident period. The relatively low redundancy, in this case, is due to the lack of alternative rail lines. Though there are some nearby bus services (see Figure \ref{fig_blueline_demand_station}), the capacity of buses is much lower than that of the metro lines. Also, most of the bus routes are not directly connected to downtown, which increases the travel time for passengers using buses. The low NURI indicates that during the Blue Line incident, CTA needs to provide more alternative services, such as dispatching shuttle buses, increasing the frequency of substitutional bus routes. 

\subsubsection{Headway analysis}\label{sec_blue_headway}
\begin{figure}[H]
\centering
\subfloat[Blue Line (southbound)]{\includegraphics[width=0.33\textwidth]{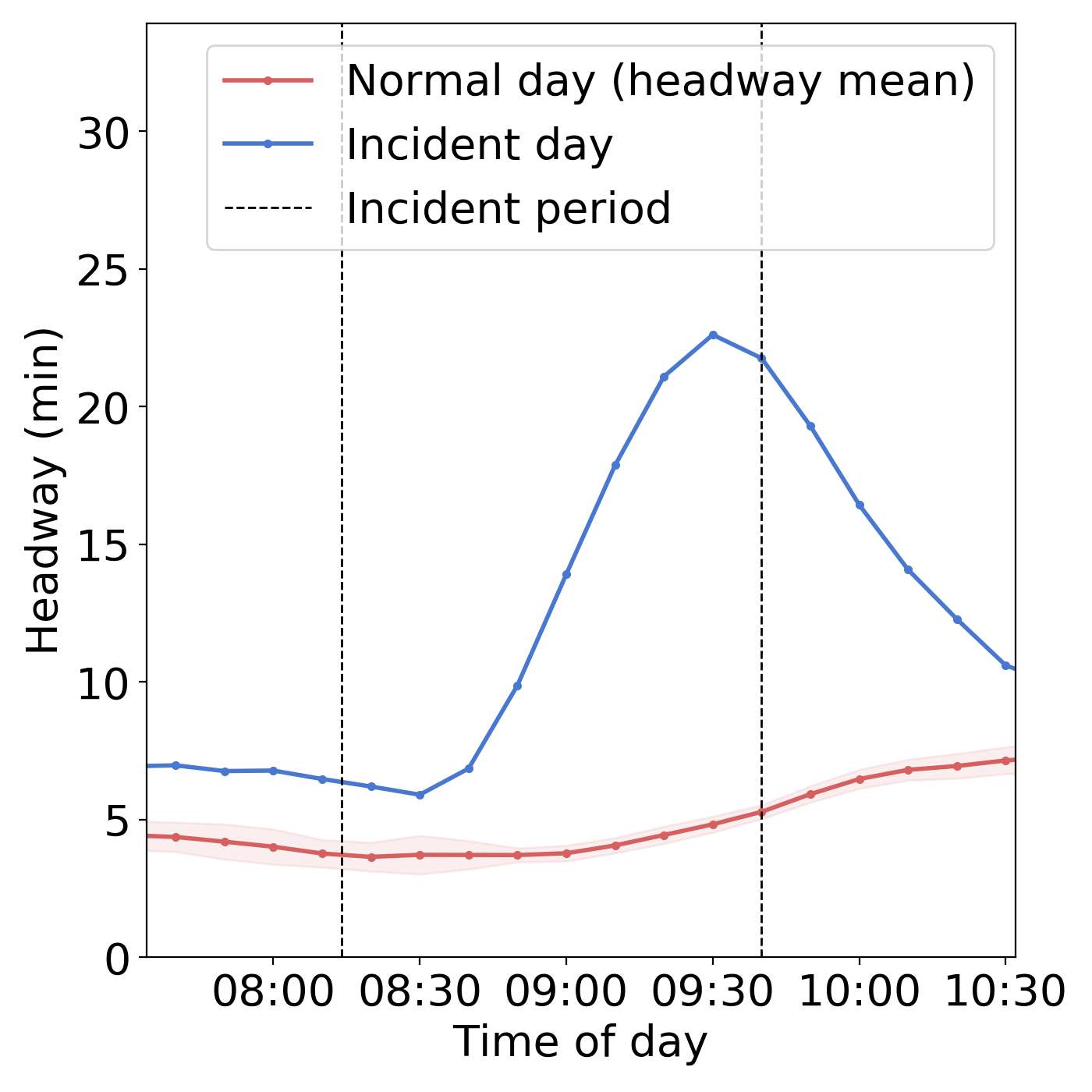}\label{fig_bluesupply_2}}
\hfil
\subfloat[Blue Line (northbound)]{\includegraphics[width=0.33\textwidth]{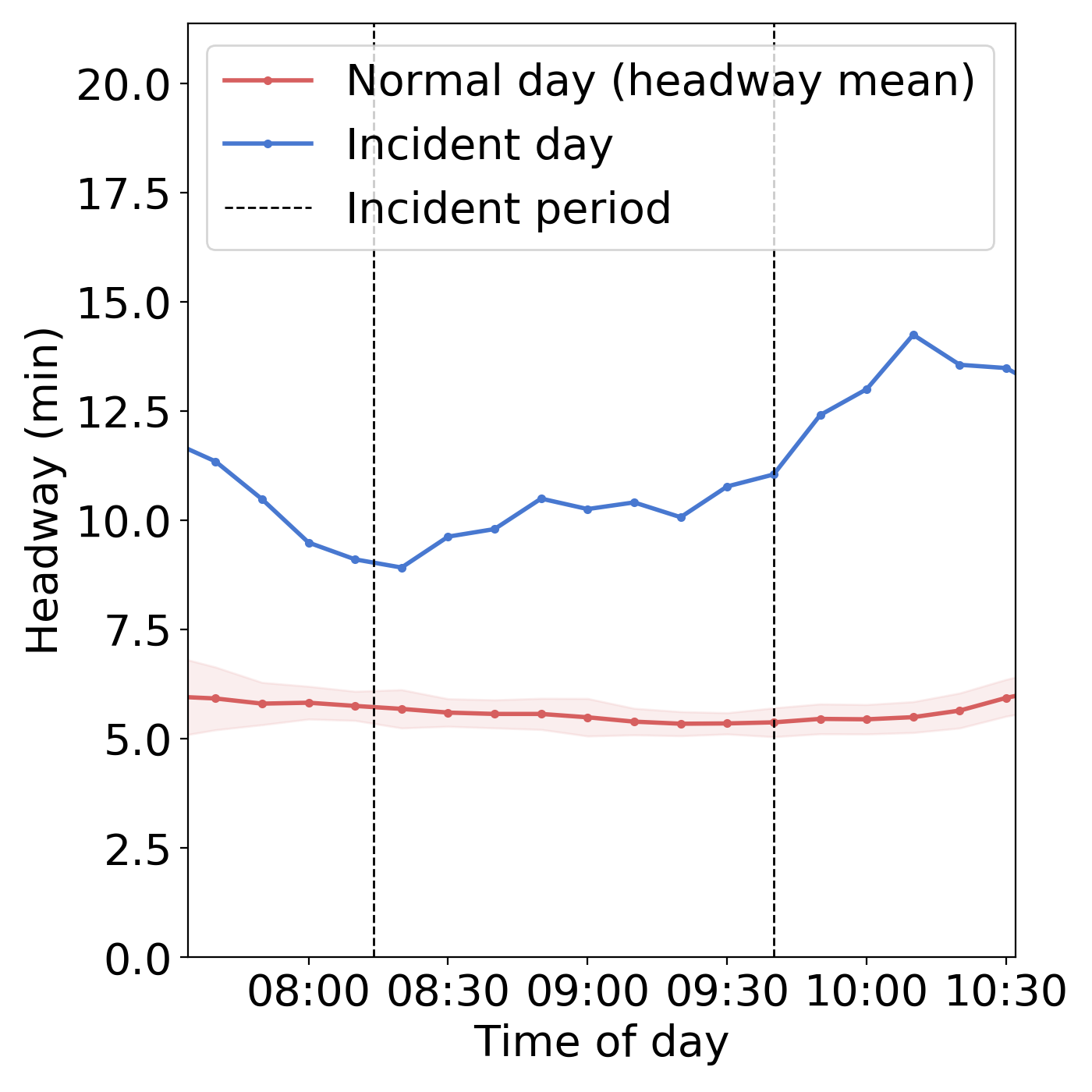}\label{fig_bluesupply_1}}
\hfil
\subfloat[Brown Line (southbound)]{\includegraphics[width=0.33\textwidth]{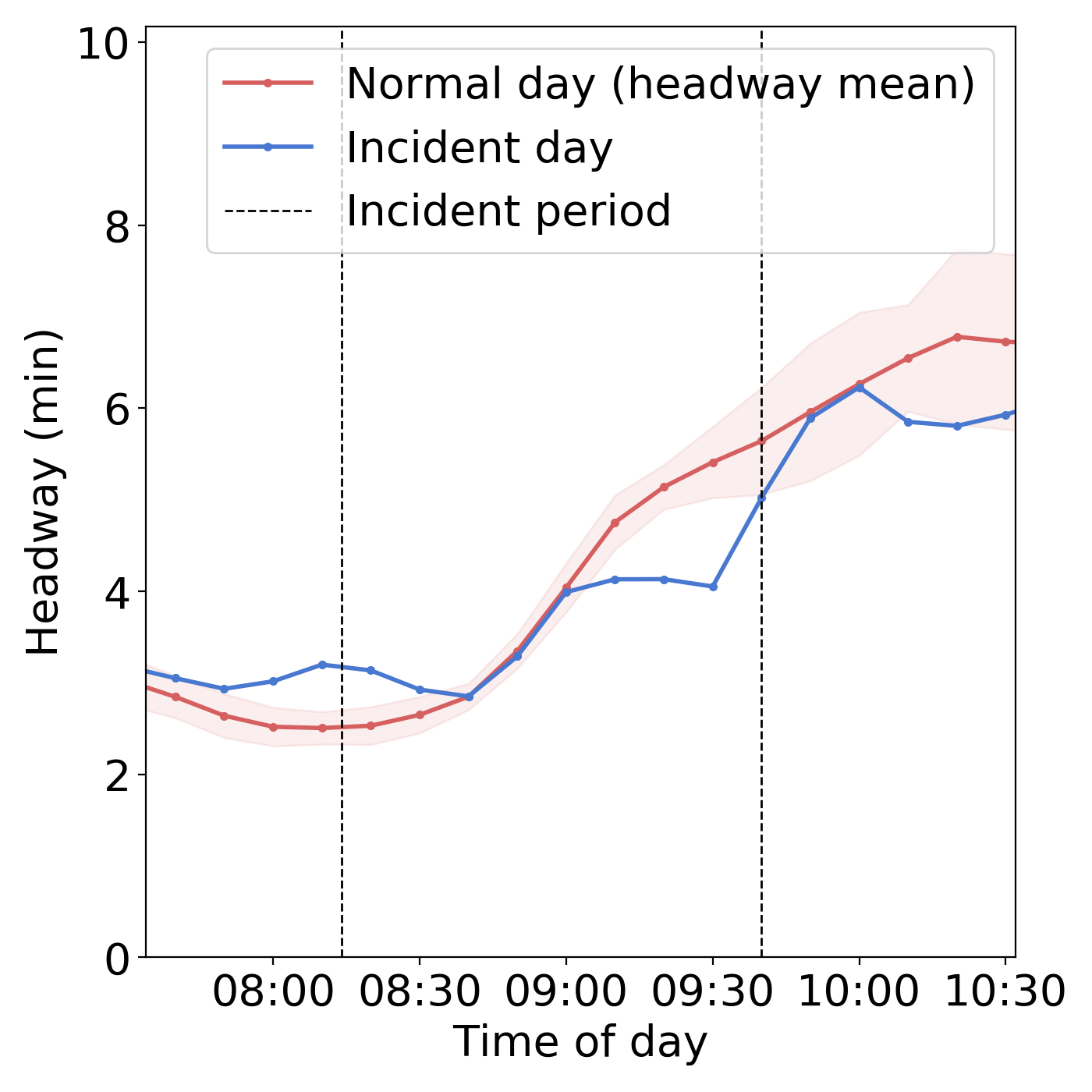}\label{fig_bluesupply_3}}
\hfil
\caption{Headway temporal distribution (Blue Line incident)}
\label{fig:supply_blueline}
\end{figure}

The headway analysis results from the Blue Line incident are shown in Figure \ref{fig:supply_blueline}. Looking at the Blue Line southbound (Figure \ref{fig_bluesupply_1}), the headway was a little bit longer than usual at the start of the day for unknown reasons. As the incident starts, the headway increases immediately for southbound trips. The increase is steeper before 9:30 AM, which is understandable since before that time CTA was working on changing the system to single-track operation. Once the single-track operation successfully deployed for all southbound trains, the headway plateaued, and then gradually decreased after 9:30 AM. On average, headway increases from 7 minutes to 17 minutes in Blue Line southbound, indicating a 58.8\% reduction in service frequency.

Figure \ref{fig_bluesupply_2} shows the headway change for the Blue Line northbound. Similarly, the headway was a little bit longer than usual at the start of the day. As the incident started, headways gradually increased. However, though the single-track operation starts at 9:30 AM, the northbound headway still remains higher than normal. This may be because CTA allowed more southbound trains to cross the single track area as they serve the major demand in the morning peak, which caused delays for the northbound trains. On average, headway increases from 8 minutes to 12 minutes in Blue Line southbound, indicating a 33.3\% reduction in service frequency.

The headway for the Brown Line southbound is also shown in Figure \ref{fig_bluesupply_3} as the Brown Line may be a possible alternative for passengers in the south part of the Blue Line. The headway remains relatively unchanged throughout the Blue Line incident, which means the incident did not affect the Brown Line operations.

% are quick to rise and continue throughout the delay despite single tracking. This is most likely due to the southbound track being the broken one. Additionally, as the blockage is close to the terminal station where trains switch from outbound to inbound, it would be fair to assume a backlog of trains clogging up the section of line from Harlem to O'Hare. Once the track is reopened, trains can run more frequently and service can resume at faster speeds, helping unclog the backlog of trains and lowering down the headway significantly. 

% the track was not opened to single tracking until after the scene was assessed. Once the track is opened to single tracking, headway plateaus but still remains higher than the average, due to both single-tracking and residual delays. Once the track is reopened to double track, headway continues to rise, most likely to now due to a backlog of trains waiting to go inbound (and thus fewer outbound trains available. We can see lasting effects for the next hour, with headways even higher than during the delay, further stressing the importance of understanding these incidents and how to best mitigate them. but remained higher than the average, due to both single-tracking and residual delays. 

\subsubsection{Passenger flow analysis}\label{sec_blue_flow}

\begin{figure}[H]
\centering
\subfloat{\includegraphics[width=0.5\textwidth]{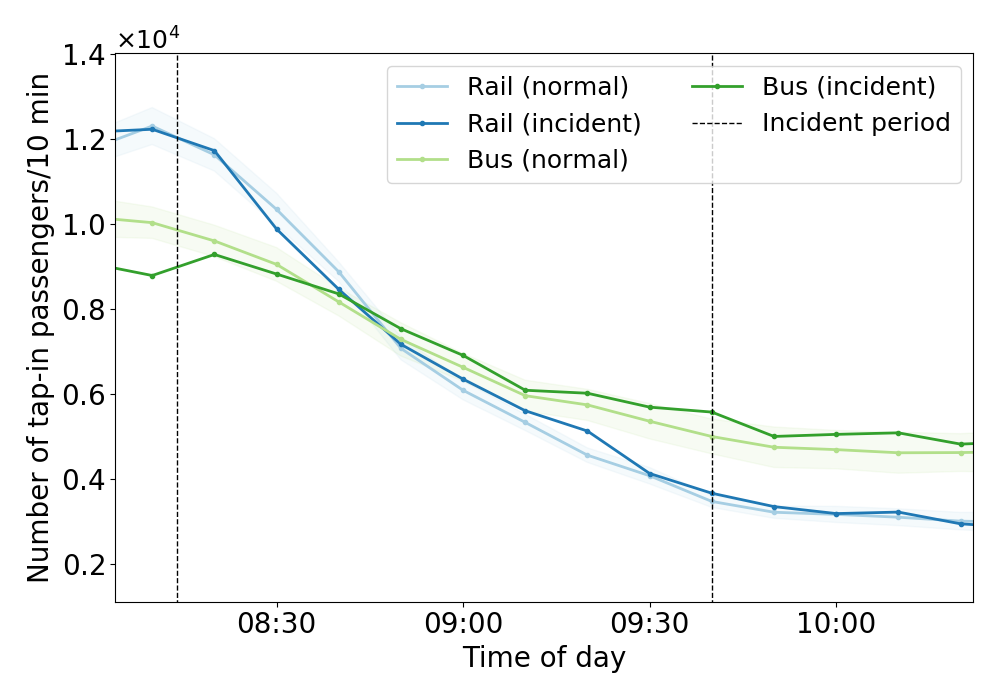}}
\caption{System level demand analysis (Blue Line incident)}
\label{fig:systemlevelblue}
\end{figure}

We first look at the system level demand change during the Blue Line incident in Figure \ref{fig:systemlevelblue}. Similar to the results from the Brown and Purple Lines incident, there is no significant difference between incident day and normal days for both bus and rail systems because the incident demand is within the 2 standard deviation range, implying that the incident did not significantly change the demand patterns for the whole system.  

\begin{figure}[H]
\centering
\subfloat[Blue Line (blocked)]{\includegraphics[width=0.33\textwidth]{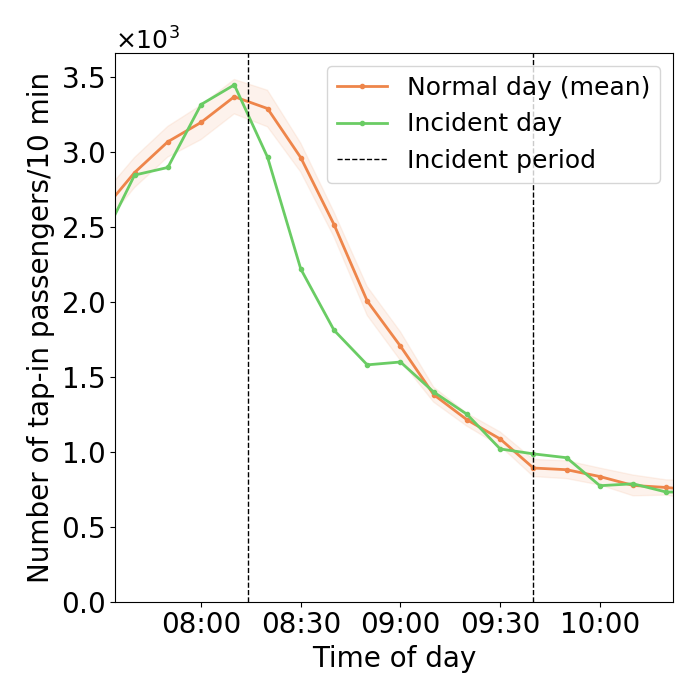}\label{fig_demandline_1_blue}}
\hfil
\subfloat[Brown Line (open)]{\includegraphics[width=0.33\textwidth]{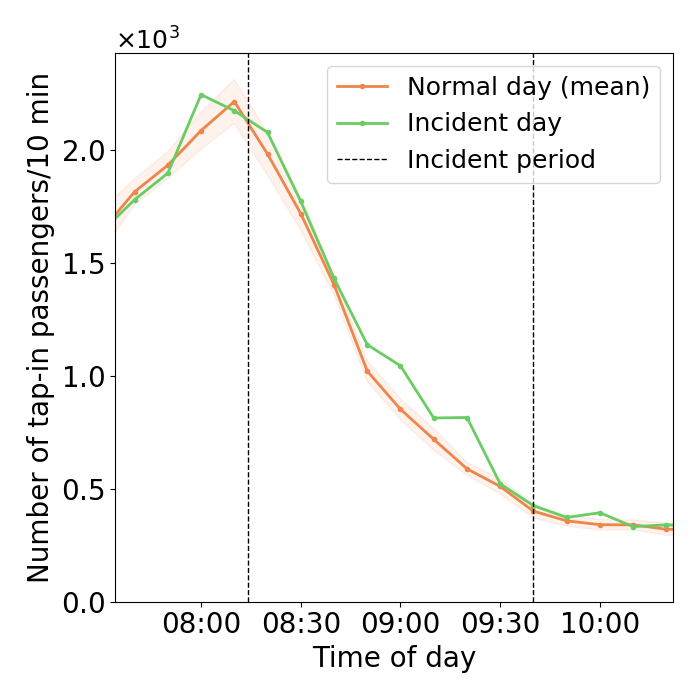}\label{fig_demandline_2_blue}}
\hfil
\subfloat[Red Line (open)]{\includegraphics[width=0.33\textwidth]{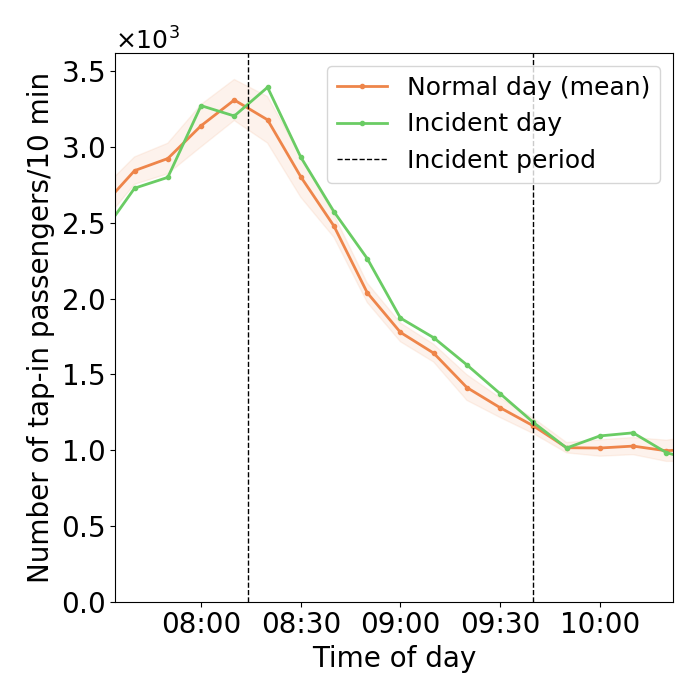}\label{fig_demandline_3_blue}}
\caption{Line level passenger flow analysis (Blue Line incident)}
\label{fig_demand_linelevel_blue}
\end{figure}

The demand patterns of the Blue, Brown, and Red Lines during the Blue Line incident are shown in Figure \ref{fig_demand_linelevel_blue}. The Blue Line (Figure \ref{fig_demandline_1_blue}) initially experiences a drop in the number of tap-in passengers immediately after the incident, which is as expected because passengers were informed of the incident and chose to not tap in. As the single-track operation started, the number of tap-ins gradually returned to regular levels as the system's backlog slowly began to clear. By 9:40 AM, single tracking is in full operation. Hence, the number of tap-in passengers is closer to average. 

For the Brown Line (Figure \ref{fig_demandline_2_blue}), we see a slight spike of demand about 30 minutes after the incident. This is because the Brown Line is not within the walking distance from the Blue Line. Passengers need to take the eastbound bus routes and then transfer onto the Brown Line to continue their journeys, which takes around 30 minutes. We also observe a consistent (though not significant) demand increase in the Red Line for the entire major incident period (Figure \ref{fig_demandline_3_blue}). The reason may be that Red and Brown Lines are largely overlapped near the incident area and can both be alternatives for the Blue Line.

\begin{figure}[H]
\centering
\subfloat[Demand changes of nearby bus stops and rail stations]{\includegraphics[width=0.9\textwidth]{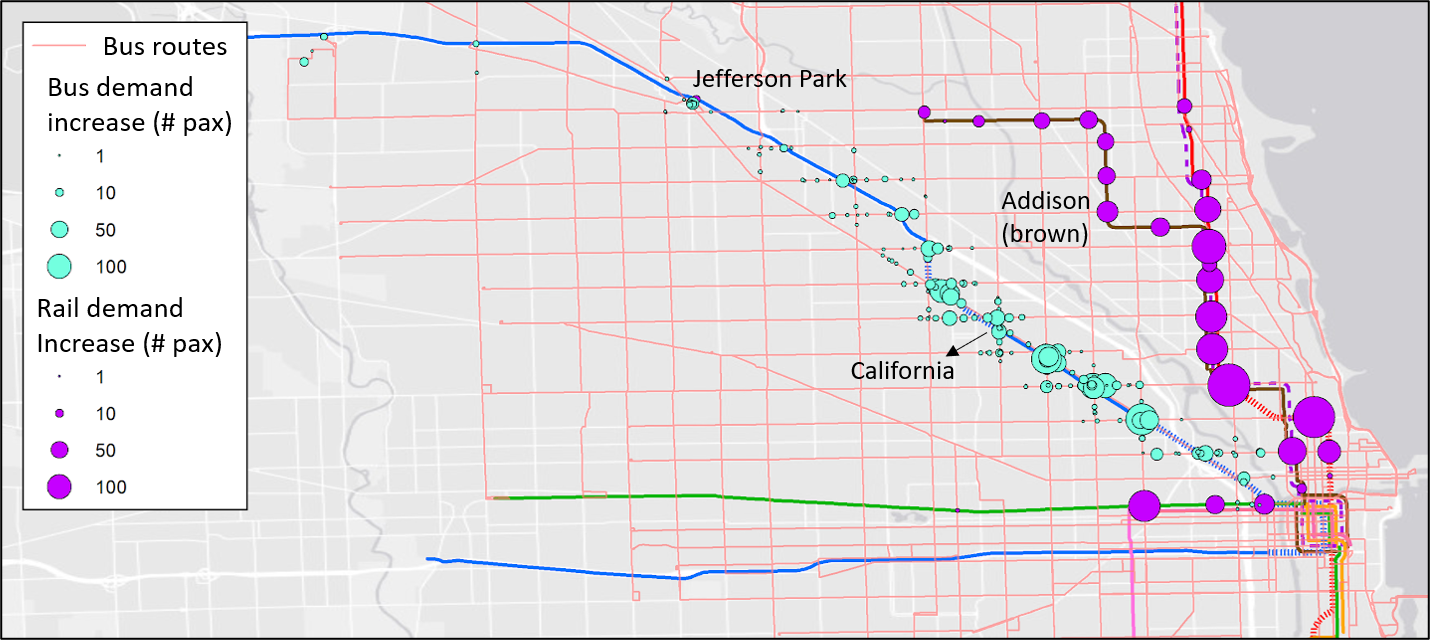}\label{fig_station_flow_1}}
\hfil
\subfloat[Demand changes of nearby bus routes]{\includegraphics[width=0.9\textwidth]{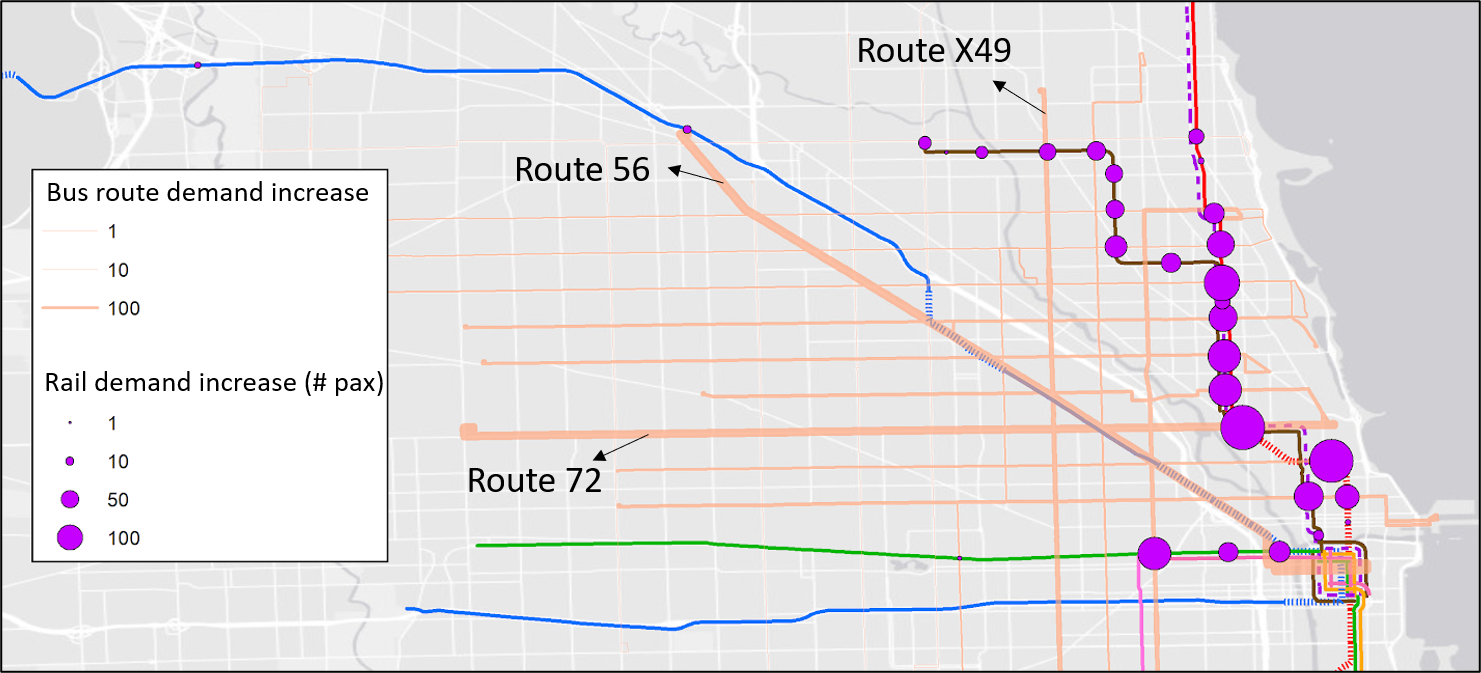}\label{fig_station_flow_2}}
\caption{Station and bus route demand increase patterns (Blue Line incident)}
\label{fig_blueline_demand_station}
\end{figure}

Demand changes at individual rail stations and bus stops near the incident area are shown in Figure \ref{fig_blueline_demand_station}. Figure \ref{fig_station_flow_1} shows that demand rises at the bus stations that are close to the Blue Line, which means many passengers switched to bus services during the incident. We also observe a substantial increase in ridership on the nearby Brown and Red Lines. This is presumably from passengers taking buses from the Blue Line and transferring to the Brown and Red Lines. However, we see little increase in ridership on the Green Line in comparison. Since the Green Line is close to the Blue Line and provides service to downtown as well, it should be a good alternative. But the small number of passengers using it implies that some passengers did not make good choices.

Figure \ref{fig_station_flow_2} illustrates the demand changes for nearby bus routes. The demand for several bus lines increased, with routes 56, 72, and X49 being the top 3. Route 56 demand increased most because it runs parallel to much of the Blue Line and connects directly to downtown. The increase in route 72 may be due to passengers transferring to that route and using it to connect to the Brown and Red Lines. Since there is little increase in the Green Line where the X49 connects, most of the increased ridership in route X49 was probably passengers with destinations in the south that route X49 directly serves. 

The total decrease of the number of tap-in passengers in the Blue Line is 2,219, while the increases in nearby bus stations, Brown Line, and Red Line are 2,426, 845, and 1,125, respectively. It is worth noting that passengers may tap in the Blue Line then get out to use buses due to long waiting times. This implies that the actual demand decrease in the Blue Line is larger than 2,219. For all passengers using nearby bus stations (2,426), most of them (845+1,125) transferred to Brown and Red Lines. 

\begin{figure}[H]
\centering
\subfloat[Jefferson Park (incident station, partially blocked)]{\includegraphics[width=0.33\textwidth]{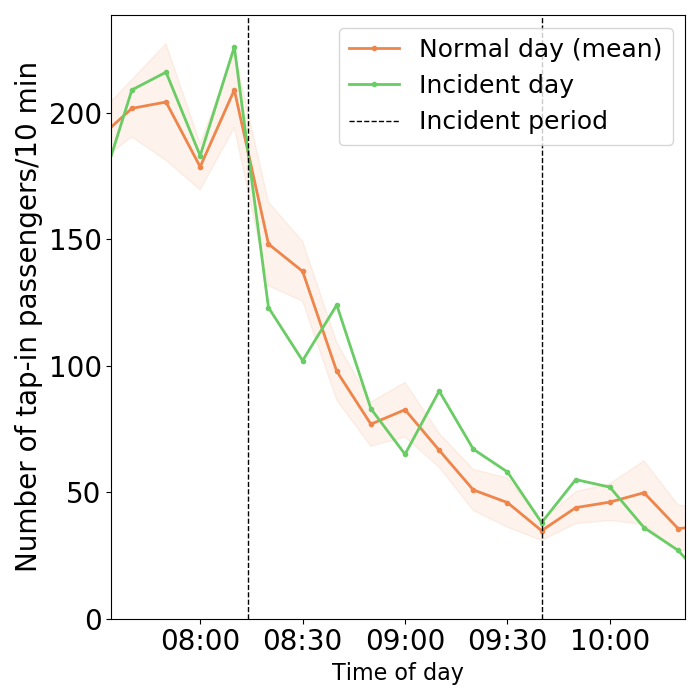}\label{JFP_JFP}}
\hfil
\subfloat[California (partially blocked)]{\includegraphics[width=0.33\textwidth]{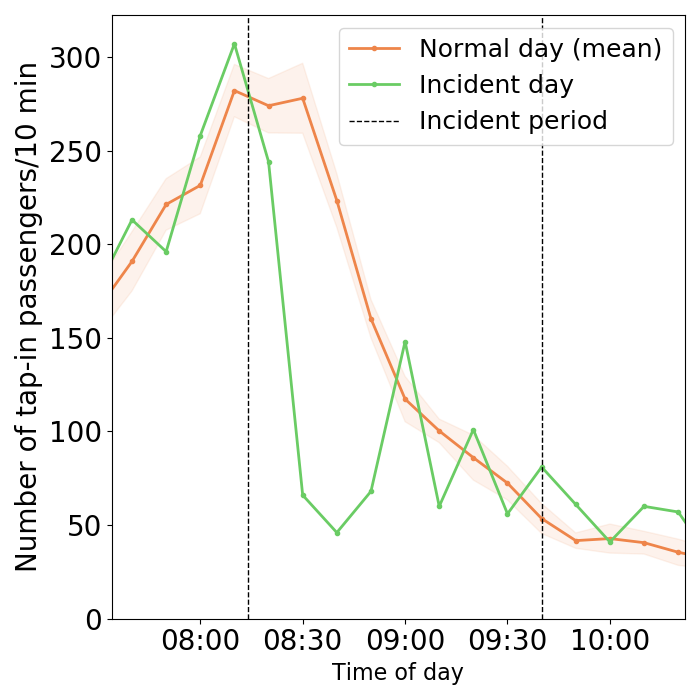}\label{JFP_California}}
\hfil
\subfloat[Addison (Brown Line, open)]{\includegraphics[width=0.33\textwidth]{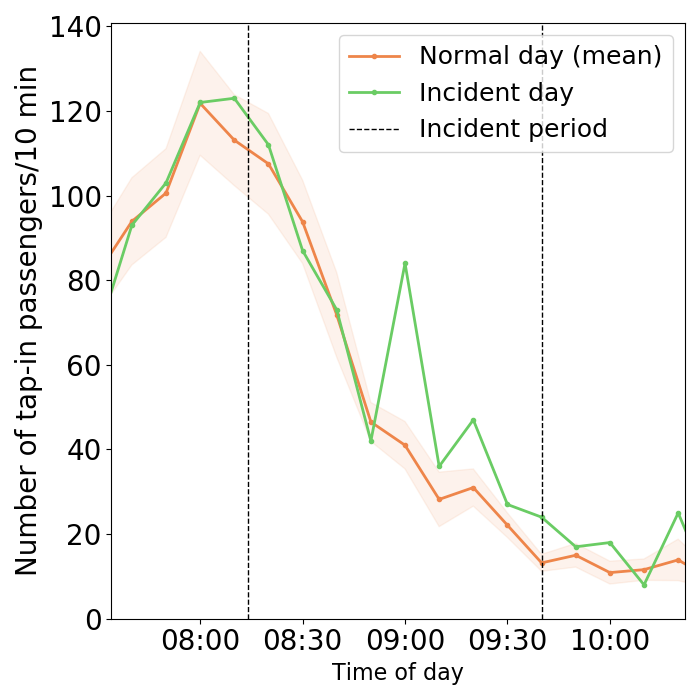}\label{JFP_Addison}}
\caption{Station level passenger flow analysis (Blue Line incident)}
\label{fig:bluelinestationanalysis}
\end{figure}

Further analysis can be done on specific key stations in terms of temporal demand patterns. Figure \ref{fig:bluelinestationanalysis} summarizes the demand changes at the Jefferson Park (incident station, partially blocked) (Figure \ref{JFP_JFP}), California, (partially blocked) (Figure \ref{JFP_California}), and Addison (Brown Line, open) (Figure \ref{JFP_Addison}) stations. At Jefferson Park, the number of tap-in passengers does not show a significant difference compared to that of normal days (i.e., within two standard deviations). Possible reasons are 1) passengers were not well informed of the incident and entered the station during the service disruption; 2) there are not enough alternative services for passengers at the Jefferson Park station, so passengers chose to enter the station and wait for service. At the California station, we see a huge drop-off in ridership before 9:00 AM. This may be due to the fact that California is closer to downtown and has more bus options for riders, which corresponds to the results in Figure \ref{fig_station_flow_1}. As the single track operations stabilize (around 9:00 AM), we see an increase in the number of tap-ins. Lastly, looking at Addison (Figure \ref{JFP_Addison}), we observe normal ridership during the first part of the incident. Halfway through, a large spike in ridership takes place. This is most likely explained by Blue Line riders taking a bus to the Brown Line, as outlined in Figure \ref{fig_station_flow_1}. And the spike is due to that they arrived as a group.

%==============================================

\subsection{Individual passenger choice analysis}\label{sec_indiv}
To analyze the individual-level passenger choices, we sampled 1,060 regular passengers who are affected by the incident (see Section \ref{sec_method_indi} for method details) using the AFC data from the two incidents above, 533 of which are from the Brown and Purple Line incident case and 527 from the Blue Line incident case. Table \ref{tab_individual} provides descriptive statistics related to various variables of interest. All transaction-related variables (such as total added value and total add-value times) are calculated based on smart card transaction data from January to December 2019. 

\begin{table}[H]
\center
\caption{Descriptive statistics of samples}
\begin{tabular}{@{}lcc@{}}
\toprule
Variables                         & Mean  & Standard deviation  \\ \midrule
Total added values (\$/year) & 917.7 & 367.8       \\
Add-value frequency (times/year)  & 31.38  & 32.76           \\
Max single added value (\$)         & 65.99  & 37.61       \\
Living in high household income area$^1$ (Yes = 1)         & 0.093  & 0.291       \\
Living in low household income area$^2$ (Yes = 1)         & 0.013  & 0.114       \\
Using pass$^3$ (Yes = 1)              & 0.374  & 0.484        \\
Reduced fare status (Yes = 1)      & 0.087  & 0.281        \\ 
OD-based redundancy  & 0.929  & 0.237        \\ 
Downtown destination (Yes = 1)      & 0.635  & 0.482        \\ 
\midrule
\multicolumn{3}{l}{Number of observations: 1,060 (533 from Brown Line case and 527 from Blue Line case)} \\
\multicolumn{3}{l}{Choices: CTA: 268; Other: 792} \\
\multicolumn{3}{l}{$^1$: Living in areas where the median annual household income is greater than \$120,000} \\
\multicolumn{3}{l}{$^2$: Living in areas where the median annual household income is less than \$25,000} \\
\multicolumn{3}{l}{$^3$: The fare type is ``pass'' on the incident day} 
\\\bottomrule
\end{tabular}
\label{tab_individual}
\end{table}

% 144 of them continued to use CTA, while 396 of them did not. Within this sample size, 495 had paid the full fare, while only 45 had reduced fare status. 226 passengers were using a pass, while 314 were using pay-to-go. The average household income was \$79,332. The average value added per time was \$52.60, with an average of 32 times per year that value was added (about every 11 days). 

The estimation results of the binary logit model are shown in Table \ref{tab_binary_logit}. ``Other'' is set as the base travel mode. We observe that passengers with larger total added value and those who use a pass (as opposed to pay-as-you-go) are more likely to choose CTA during the incident. This is understandable because these passengers generally use the public transit system more frequently. They are familiar with the service and able to find alternative public transit routes during the incident. Passengers who live in high household income areas and have high max single added value are less likely to choose CTA (both are significant at 0.15 level). Note that both of these two variables are used as proxies for the high income. Hence, their choice of other options may be because they can afford alternative modes of transportation (such as Uber/Lyft). Passengers with reduced fare status are more likely to use CTA services. The reason may be that reduced fare status users are usually students, seniors, and disabled people likely on limited incomes. They usually rely primarily on CTA to travel. OD-based redundancy has a positive impact on choosing CTA, which is as expected because higher redundancy indicates better alternative public transit services. Another interesting result is that passengers with the destinations in the downtown area are less likely to use CTA. This may be because these passengers were going to work and they have a higher motivation to arrive on time, thus changing to alternative modes (such as Uber/Lyft).

\begin{table}[H]
\center
\caption{Individual choice model estimation results}
\begin{tabular}{@{}lcl@{}}
\toprule
Parameters                                                 & Value (standard error)                    &                        \\ \midrule
CTA: ASC                         & -3.27 (0.522)                            & $^{***}$                      \\
CTA: Total added value (\$1000/year)                          & 1.26 (0.277)                            &  $^{***}$                      \\
CTA: Add-value frequency (100 times/year)                         & -0.449 (0.365)                            &                        \\
CTA: Max single added value (\$1000)                         & -5.89 (3.72)                            & $\cdot$                       \\
CTA: Living in high household income area (Yes = 1)                          & -0.396 (0.269)                          & $\cdot$                   \\
CTA: Living in low household income area (Yes = 1)                          & 1.380 (0.568)                          & $^{**}$                   \\
CTA: Using pass (Yes = 1)                                  & 1.13 (0.215)                              & $^{***}$                    \\
CTA: Reduced fare status (Yes = 1)                         & 0.627 (0.297)                             & $^{**}$                      \\
CTA: OD-based redundancy                         & 1.26 (0.432)                             & $^{***}$                      \\
CTA: Downtown destination (Yes = 1)                         & -0.335 (0.159)                             & $^{**}$                      \\
Other: ASC                     & 0 (fixed)                                 &                        \\ \midrule
\multicolumn{3}{l}{\begin{tabular}[c]{@{}l@{}}Number of individuals: 1060. Adjusted $\rho^2 = 0.245$\\ $^{***}$: $p<0.01$; $^{**}$: $p<0.05$; $^{*}$: $p<0.1$; $\cdot$: $p<0.15$  \end{tabular}}
\end{tabular}
\label{tab_binary_logit}
\end{table}

\begin{figure}[H]
\centering
\subfloat{\includegraphics[width=0.6\textwidth]{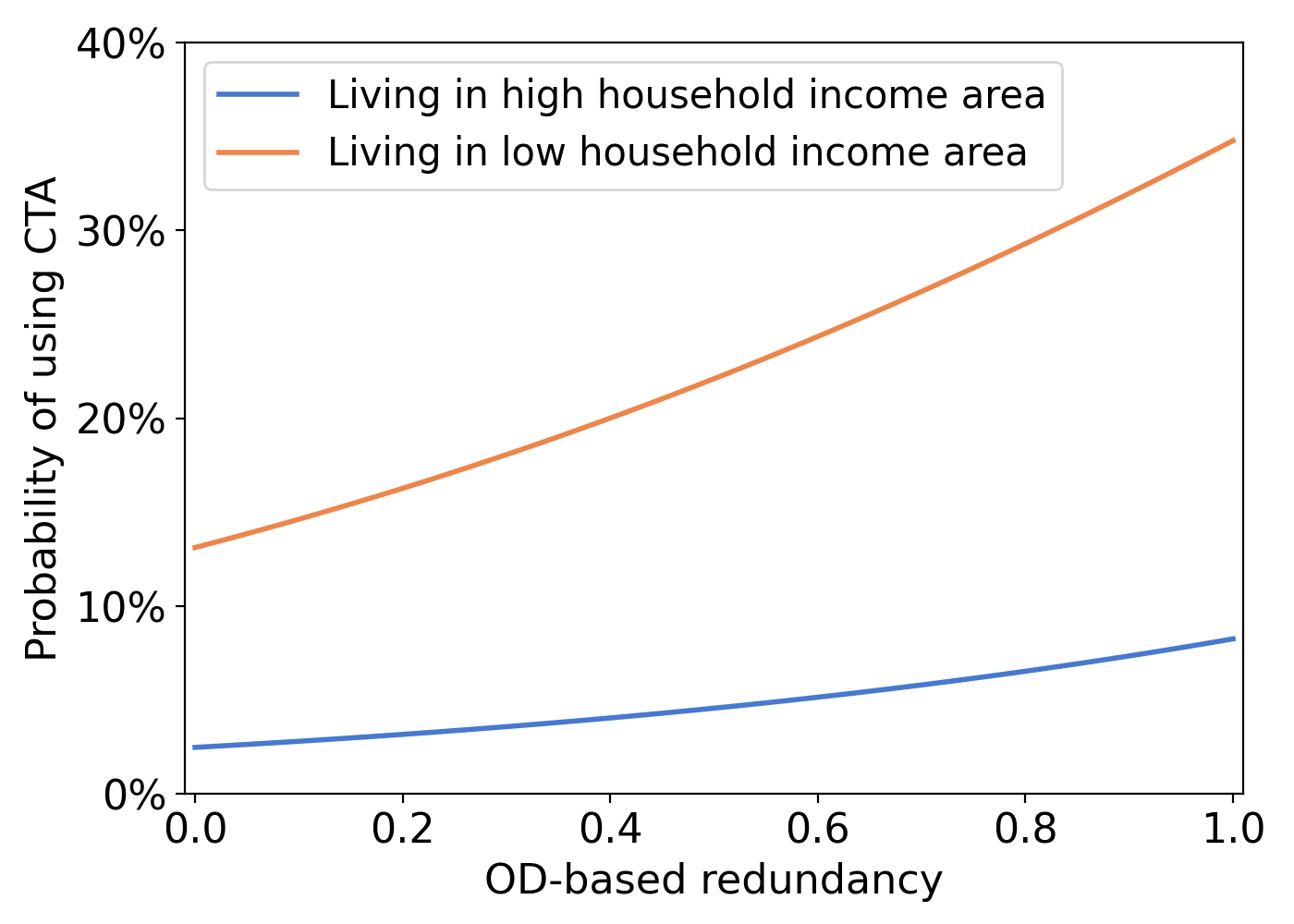}}
\caption{Impact of OD-based redundancy for passengers living in high and low income areas}
\label{fig_impact_redundancy}
\end{figure}

We also evaluate the sensitivity of the probability of choosing CTA with respect to the OD-based redundancy (Figure \ref{fig_impact_redundancy}). The probabilities in Figure \ref{fig_impact_redundancy} are calculated by fixing the remaining variables to the corresponding sample means. Similar to the results above, low-income passengers have a higher probability of using CTA than that of high income, and the difference increases with the increase in redundancy. This implies that low-income passengers have higher elasticity with respect to redundancy. Assuming OD-based redundancy equal to 0.5, a 1\% increase in OD-based redundancy can lead to a 0.03\% increase in the probability of choosing CTA for high-income passengers, and a 0.11\% increase for low-income passengers.

Understanding the impact of demographics on travel mode choices is helpful for transit operators to customize their operation strategies during the incident. For example, as low-income passengers are more likely to use CTA during the incident, alternative services can be provided to serve low-income areas first. 

\section{Conclusion and Discussion}\label{sec_conclude}
\subsection{Conclusion}
This study proposes a general incident analysis framework both from the supply and demand sides using automatically collected data (AFC and AVL) in public transit systems. Specifically, from the supply side, we propose an incident-based network redundancy index to analyze the network’s ability to provide alternative services under a specific rail disruption. The impacts on service operations are analyzed through the headway changes. From the demand side, we calculate the demand changes at different rail lines, rail stations, bus routes, and bus stops to understand the passenger flow redistribution under incidents. Individual behavior is analyzed using a binary logit model based on inferred passengers’ mode choices and socio-demographics inferred from AFC and sale transaction data. Two incidents in the CTA public transit system are used as case studies. The two rail disruption cases have different attributes, one at a location with high network redundancy and the other with low network redundancy.  

Results show that the service frequency of the incident line was largely reduced during the incident time. Nearby lines with substitutional functions are also slightly affected. Depending on the incident location, the network's redundancies are different, as well as the passengers' behavior. In the low redundancy scenario, most of the passengers chose to use nearby buses to move, either to their destinations or to the nearby rail lines. In the high redundancy scenario, most of the passengers transferred directly to nearby rail lines.  

\subsection{Policy implications and suggestions}
The results of the case study provide useful insights into operations when dealing with incidents. We summarize the main policy implications below.

\textbf{Planing for incident responses using redundancy index}. In Section \ref{sec_redun}, we calculate the NRUI for different stations by assuming a one-hour track-block incident. The NRUI can be adapted to different types of incidents, network blockages, and duration. Based on a graph similar to Figure \ref{fig_cta_redundancy}, transit operators can better plan for future incidents, such as planning alternatives services for low-redundancy locations, preparing route recommendation strategies for high-redundancy locations, etc. 

\textbf{Headway management for both the incident line and nearby lines}. In Sections \ref{sec_brown_headway} and \ref{sec_blue_headway}, we observe that headways increase in both the incident line and nearby lines. The results suggests that transferred passengers from the incident line and unusual operations of the incident line may also affect operations of nearby lines. More comprehensive headway management should be considered during incidents.

\textbf{Provision of timely customer information}. The results indicate that passengers tap into the blocked station during the incident, implying that these passengers are not well informed. Transit agencies should improve their customer information delivery during incidents (especially at fare gates). This can be done through text messaging, Twitter, in-station signs, station staff, and a variety of other methods to keep the passenger informed. 

\textbf{Provision of route recommendations during incidents}. During the Blue Line incident, not many passengers use the Green Line, although it is a good alternative (see Section \ref{sec_blue_flow}). This suggests that passengers may not act rationally, or they lack knowledge about the available alternatives. Providing route recommendations to passengers during the incident can increase utilization of alternative services and improve level of service. 

\textbf{Data-driven methods to design alternative services}. The analysis provides a better understanding of how passengers move and the alternatives they may choose, based on which operators can better allocate available buses or trains. For example, most passengers used Bus Route 56 as a substitutional service during the Blue Line incident (Section \ref{sec_blue_flow}). CTA may increase the service frequency of these heavily used routes.

\textbf{Provision of shuttle services to improve use of alternative routes}. During the Blue Line incident, one of the reasons that the Green Line is not fully utilized may be that it is not directly connected to the Blue Line (Section \ref{sec_blue_flow}). Hence, CTA may provide shuttle services to connect the Blue and Green lines to encourage more passengers to follow the recommendation (note that multiple recommendations should be provided to avoid overwhelming of a specific line). 

% \textbf{Operation adjustment based on demographics}. As shown in Section \ref{sec_indiv}, passengers with different demographics may respond differently to the incident. Specifically, low-income passengers and passengers with reduced fare status more rely on public transit during incidents. Hence, transit agencies may provide more alternative services to low-income areas and if possible, provide free rides to reduced-fare-status users. For high-income areas, transit authorities may cooperate with transportation network companies (TNC, such as Uber/Lyft) and ask them to provide more vehicles in high-income areas (note that traffic congestion should be considered when TNC's supply is increased). 

\section{Acknowledgements}
The authors would like to thank the Chicago Transit Authority (CTA) for their support and data availability for this research. The authors acknowledge that the initial version of this paper has been presented in the Transportation Research Board 100th Annual Meeting.

\section{Declaration}

% \subsection{Confict of interest}
On behalf of all authors, the corresponding author states that there is no conflict of interest.

\section{Authors’ contribution}
\textbf{Baichuan Mo}: Conceptualization, Methodology, Software, Formal analysis, Data Curation, Writing - Original Draft, Writing - Review \& Editing, Visualization. \textbf{Max Y von Franque}: Software, Formal analysis, Writing - Original Draft, Visualization. \textbf{Haris N. Koutsopoulos}: Conceptualization, Formal analysis, Writing - Review \& Editing, Supervision. \textbf{John Attanucci}: Conceptualization, Supervision, Project administration, Funding acquisition. \textbf{Jinhua Zhao:} Conceptualization, Supervision, Project administration, Funding acquisition.

\bibliography{mybibfile}

\appendix
\appendixpage
\section{Proof of $\lim_{D_I \to \infty} A_{p}$} \label{append_cap}
Let $k^*$ be the last trip of path $p$ that can reach the destination during the incident period. Mathematically, $k^* = \argmin_k \{k =1,2,..., \floor{D_I/H_p} \;|\; D_I - (k-1)H_p \leq L_p \}$. Therefore, we have
\begin{align}
\min{\{D_I - (k-1)H_p, L_p\}} &= L_p \quad\quad\; \forall k \leq k^*\\
\min{\{D_I - (k-1)H_p, L_p\}} &= D_I - (k-1)H_p \quad\quad\; \forall k^* < k \leq \floor{D_I/H_p}
\end{align}
This leads to
\begin{align}
\lim_{D_I \to \infty} A_{p} &= \lim_{D_I \to \infty} \frac{1}{D_I} \sum_{k=1}^{\floor{D_I/H_p}} \frac{\min{\{D_I - (k-1)H_p, L_p\}}}{L_p } \cdot C_p \nonumber \\
&=\lim_{D_I \to \infty} \sum_{k=1}^{k^*} \frac{L_p}{L_p \cdot D_I} \cdot C_p + \lim_{D_I \to \infty} \sum_{k=k^*+1}^{\floor{D_I/H_p}} \frac{D_I - (k-1)H_p}{L_p \cdot D_I} \cdot C_p \\
&=\lim_{D_I \to \infty} k^* \frac{1}{D_I} \cdot C_p + \lim_{D_I \to \infty} \sum_{k=k^*+1}^{\floor{D_I/H_p}} \frac{D_I - (k-1)H_p}{L_p \cdot D_I} \cdot C_p \nonumber
\end{align}
Notice that $k^*\to \floor{D_I/H_p}$ as $D_I\to \infty$ because when $D_I$ is large enough, almost all trips can reach the destination. And $\lim_{D_I\to \infty} \floor{D_I/H_p} =  {D_I/H_p}$ by definition. Therefore,
\begin{align}
\lim_{D_I \to \infty} A_{p} &= \lim_{D_I \to \infty} \floor{D_I/H_p} \cdot \frac{1}{ D_I} \cdot C_p + 0 = \frac{C_p}{H_p}
\end{align}

\end{document}